\documentclass[10pt,journal,compsoc]{IEEEtran}

\ifCLASSOPTIONcompsoc
  \usepackage[noadjust]{cite}
\else
  \usepackage{cite}
\fi

\ifCLASSINFOpdf
   \usepackage[pdftex]{graphicx}
   \graphicspath{{img/}}
\else
\fi

\usepackage{subcaption}

\usepackage{amsmath}
\DeclareMathOperator*{\argmax}{argmax} 
\usepackage{algorithmic}
\usepackage{array}
\usepackage{url}
\usepackage{multirow, booktabs}
\usepackage{enumitem}
\usepackage{amsmath,amsfonts,amssymb}
\usepackage{pgfplotstable}
\pgfplotsset{compat=1.12}
\usetikzlibrary{patterns}
\usepgfplotslibrary{groupplots}

\usepackage{soul}
\usepackage{makecell} 
\usepackage{xcolor}
\definecolor{githubblue}{HTML}{0366d6}
\usepackage{hyperref}
\hypersetup{
    colorlinks=true,    
    urlcolor=githubblue,       
}
\usepackage{xurl}
\usepackage{rotating}
\usepackage{listings}
\usepackage{tcolorbox}
\usepackage{fontawesome}
\newcommand{\blue}[1]{\textcolor{black}{#1}}
\newcommand{\red}[1]{\textcolor{black}{#1}}

\begin{document}
\title{AutoAlign: Fully Automatic and Effective Knowledge Graph Alignment enabled by Large Language Models}
\author{Rui~Zhang$^\dag$,
        Yixin~Su$^\dag$\faEnvelopeO,
        Bayu~Distiawan~Trisedya,
        Xiaoyan~Zhao,
        Min~Yang,
        Hong~Cheng,
        and~Jianzhong~Qi
\IEEEcompsocitemizethanks{
    \IEEEcompsocthanksitem R. Zhang is with Tsinghua University. E-mail: rayteam@yeah.net.
    \IEEEcompsocthanksitem Y. Su is with The University of Melbourne. E-mail: yixin.su@outlook.com.
    \IEEEcompsocthanksitem B.D. Trisedya is with Universitas Indonesia. E-mail: b.distiawan@cs.ui.ac.id.
    \IEEEcompsocthanksitem X. Zhao, and H. Cheng are with The Chinese University of Hong Kong, Hong Kong, China. E-mail: \{xzhao, hcheng\}@se.cuhk.edu.hk.
    \IEEEcompsocthanksitem M. Yang is with the Shenzhen Institute of Advanced Technology, Chinese Academy of Sciences, Shenzhen, China. E-mail: min.yang@siat.ac.cn.
	\IEEEcompsocthanksitem J. Qi is with The University of Melbourne. E-mail: jianzhong.qi@unimelb.edu.au.
}
\thanks{$^\dag$~R. Zhang and Y. Su are co-first authors.}
\thanks{\faEnvelopeO ~Y.~Su is the corresponding author.}
\thanks{Manuscript received XXXXXX XX, XXXX; revised XXXXXX XX, XXXX.}
}


\IEEEtitleabstractindextext{%
 
\begin{abstract}\label{abstract}
	The task of entity alignment between knowledge graphs (KGs) aims to identify every pair of entities from two different KGs that represent the same entity. 
	Many machine learning-based methods have been proposed for this task. However, to our best knowledge, existing methods all require \emph{manually crafted} seed alignments, which are expensive to obtain. 
	In this paper, we propose the first fully automatic alignment method named AutoAlign, which does not require any manually crafted seed alignments. 
	Specifically, for predicate embeddings, AutoAlign constructs a predicate-proximity-graph \red{with the help of large language models} to automatically capture the similarity between predicates across two KGs. For entity embeddings, AutoAlign first computes the entity embeddings of each KG independently using TransE, and then shifts the two KGs' entity embeddings into the same vector space by computing the similarity between entities based on their attributes. Thus, both predicate alignment and entity alignment can be done without manually crafted seed alignments. AutoAlign is not only fully automatic, but also highly effective.
	Experiments using real-world KGs show that AutoAlign improves the performance of entity alignment significantly compared to state-of-the-art methods. 
        Our source code is available at \href{https://github.com/ruizhang-ai/AutoAlign}{github.com/ruizhang-ai/AutoAlign}.
\end{abstract}
}

\maketitle
\IEEEdisplaynontitleabstractindextext
\IEEEpeerreviewmaketitle

\IEEEraisesectionheading{
\section{Introduction} 
	\label{kba-intro}}
Knowledge bases in the form of \emph{knowledge graphs} (KGs) have been used in many applications, including question answering systems~\cite{wu2017image}, dialogue systems~\cite{yang2021unimf}, and recommender systems ~\cite{zhang2016rec}. Many KGs have been created separately for particular purposes. The same real-world entity may exist in different forms in different KGs. For example, a village named \texttt{Kromsdorf} in Germany is a real-world entity that exists in two different KGs, LinkedGeoData \cite{stadler2012linkedgeodata} and DBpedia \cite{auer2007dbpedia}. This entity is denoted in the form of \texttt{lgd:240111203} in LinkedGeoData but in the form of \texttt{dbp:Kromsdorf} in DBpedia. Usually, these KGs complement each other in terms of the number of entities each KG contains, and the types of information related to each entity. Therefore, we may merge two KGs into one with more entities and richer information related to each entity. To merge two KGs, a core task is \emph{entity alignment}, which is to identify every pair of entities from the two KGs that correspond to the same real-world entity. Existing methods require significant manual work (e.g., manually crafted seed alignments), and the performance of the alignment is low. In this paper, we propose a novel method to this problem, which is fully automatic and effective (i.e., the alignment result is of high accuracy).

\begin{table}[t!]
	\centering
	\caption{Knowledge graph alignment example.}
	\begin{small}
		\begin{tabular}{@{}l@{}}
			\toprule[2pt]
			\midrule
			\multicolumn{1}{c}{$\mathcal{G}_1$} \\
			\midrule
			$\langle$\texttt{lgd:240111203,lgd:population,1595}$\rangle$ \\
			$\langle$\texttt{lgd:240111203,rdfs:label,"Kromsdorf"}$\rangle$ \\
			$\langle$\texttt{lgd:240111203,geo:lat,50.9988888889}$\rangle$ \\
			$\langle$\texttt{lgd:240111203,lgd:alderman,"B. Grobe"}$\rangle$ \\
			$\langle$\texttt{lgd:240111203,lgd:is\_in,lgd:51477}$\rangle$ \\
			\midrule
			\multicolumn{1}{c}{$\mathcal{G}_2$} \\
			\midrule
			$\langle$\texttt{dbp:Kromsdorf,rdfs:label,"Kromsdorf"}$\rangle$ \\
			$\langle$\texttt{dbp:Kromsdorf,geo:lat,50.9989}$\rangle$ \\
			$\langle$\texttt{dbp:Kromsdorf,dbp:populationTotal,1595}$\rangle$ \\
			$\langle$\texttt{dbp:Kromsdorf,dbp:located\_in,dbp:Germany}$\rangle$ \\
			$\langle$\texttt{dbp:Kromsdorf,dbp:district,dbp:Weimarer}$\rangle$ \\
			\midrule
			\multicolumn{1}{c}{Merged $\mathcal{G}_{M}$} \\
			\midrule
			$\langle$\texttt{lgd:240111203,:population,1595}$\rangle$ \\
			$\langle$\texttt{lgd:240111203,:label,"Kromsdorf"}$\rangle$ \\
			$\langle$\texttt{lgd:240111203,:lat,50.9988888889}$\rangle$ \\
			$\langle$\texttt{lgd:240111203,:alderman,"B. Grobe"}$\rangle$ \\
			$\langle$\texttt{lgd:240111203,:is\_in,lgd:51477}$\rangle$ \\
			$\langle$\texttt{lgd:240111203,:district,dbp:Weimarer}$\rangle$ \\
			\midrule
			\bottomrule[2pt]
		\end{tabular}%
	\end{small}
	\label{table-kba-example}%
\end{table}%

We use an example as shown in Table~\ref{table-kba-example} to illustrate the entity alignment problem in detail. Typically, knowledge or real-world facts in KGs are stored in the form of triples, 
and a triple consists of three elements in the form of  $\langle \textit{head}, \textit{predicate}, \textit{tail} \rangle$, where \textit{head} denotes an entity and \textit{tail} denotes either another entity or a literal (attribute value) of the head entity. Here, if \textit{tail} is an entity, the triple is called a \emph{relation triple} and the predicate is called \emph{relation predicate}; if \textit{tail} is a literal, the triple is called an \emph{attribute triple} and the predicate is called \emph{attribute predicate}. Table~\ref{table-kba-example} gives an example of two subsets of triples from two KGs, denoted by $\mathcal{G}_1$ and $\mathcal{G}_2$ (we use prefixes \texttt{lgd:} and \texttt{dbp:} to simplify the original spell out). The head entities in these two subsets refer to the same entity \texttt{Kromsdorf}, even though they are in different forms, \texttt{lgd:240111203} and \texttt{dbp:Kromsdorf}. We aim to identify such entities and give them a unified ID such that both KGs can be merged together through them. In Table~\ref{table-kba-example}, $\mathcal{G}_{M}$ denotes the merged KG with entities aligned, where \texttt{lgd:240111203} is used as the unified ID for the entity \texttt{Kromsdorf} which has a set of properties that is the union of the sets of properties from both KGs.

\begin{figure}[t!]
	\begin{center}
		\includegraphics[width=.45\textwidth]{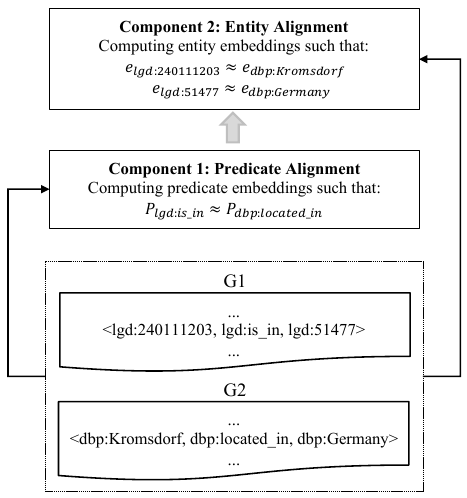}
	\end{center}
	\caption{\label{fig-kba-components} Components of knowledge graph alignment: 
\emph{predicate alignment} and \emph{entity alignment}. }
	\vspace{-3mm}
\end{figure}%

As illustrated by the above example, to align entities, we also need to have the corresponding predicates aligned (e.g., \texttt{lgd:is\_in} and \texttt{dbp:located\_in}). The task of \emph{knowledge graph alignment} is to have both entity alignment and predicate alignment between two KGs.
Recent KG alignment methods are mainly based on representation learning \cite{zhang2022benchmark}. 
Fig~\ref{fig-kba-components} shows the two critical components, 
\emph{predicate alignment} and \emph{entity alignment}:
(i) the embeddings of predicates that represent the same relationship in the two KGs should have similar embeddings in the aligned vector space, e.g., \texttt{lgd:is\_in} and \texttt{dbp:located\_in} should have close embeddings, and (ii) if entity $e_{g_1}$ from $\mathcal{G}_1$ corresponds to the same real-world entity as entity $e_{g_2}$ from $\mathcal{G}_2$, then $e_{g_1}$ should have similar embeddings to that of $e_{g_2}$ in the aligned vector space, e.g., \texttt{lgd:240111203} and \texttt{dbp:Kromsdorf} in Fig.~\ref{fig-kba-components}  should have similar embeddings. 

There are mainly two paradigms of KG
embedding, \emph{translation-based methods} and
\emph{Graph Neural Network (GNN)-based methods}. See \cite{zhang2022benchmark} for a comprehensive survey.
Translation-based methods \cite{chen2017multilingual,chen2017multigraph,zhu2017iterative} learn an embedding space for each KG separately, then learn a \emph{transition matrix} to map the embedding space from one KG to the other. The mapping relies on large numbers of seed alignments (i.e., a set of manually crafted aligned triples from the two KGs) to compute the transition matrix.  
The other paradigm, GNN-based methods \cite{cao2019multi,sun2020knowledge,wu2019relation}, aggregates information from entities' neighborhoods with the graph structure to compute entity embeddings. Then they align the space of the two KGs via manually crafted seed alignments, which is similar to translation-based methods. 
Moreover, all the existing studies have focused only on entity alignment, whereas for predicate alignment, they also rely on manually crafted seeds. In summary, to our best knowledge, existing methods of both paradigms rely on manually crafted seed alignments.

Relying on manually created seed alignments have significant drawbacks: 1) Manually created seeds require careful human curation and usually domain expertise, which is expensive. For large datasets, a substantial number of manual alignments are required, which is prohibitive. 2) The portability of manually created seeds is poor. For each new alignment task, we have to manually create seeds again. 3) Different annotators have different biases and manually created seeds are error-prone, which results in manual seeds of highly varying quality and hence uncertain quality of alignment results.

To address the above problems, we propose a novel method for KG alignment that is not only \emph{fully automatic} (i.e., not involving any manual seed alignments) but also much more accurate in aligning entities and predicates (i.e., more effective).  
We name our method AutoAlign as it is
an automatic KG alignment method without needing the human annotation of seed alignments.
For predicate alignment, AutoAlign constructs a predicate-proximity-graph to automatically capture the similarity between predicates across two KGs by learning the attentions of entity types.
The predicate-proximity-graph construction is made automatic by leveraging recent large language models (such as ChatGPT and Claude) for aligning entity types of two KGs. \blue{For entity alignment}, AutoAlign computes the entity embeddings of each KG independently using TransE, and then shifts the two KGs’ entity embeddings into the same vector space by computing the similarity between entities based on their attributes.
The learning process of the above predicate alignment and entity alignment are jointly performed, which yields the final aligned KG.

Specifically, to achieve predicate embedding alignment without manually crafted seed alignments, we propose a \emph{predicate-proximity-graph} for approximately computing the predicate embeddings, including both relation predicates and attribute predicates, where each predicate is a vertex that represents a relationship between entity types or literal types (instead of entities or literals). We create such a graph by replacing the head entity and tail entity of KG triples by their corresponding types, which are provided as \texttt{rdfs:type} relationship in the knowledge graph. For example, we replace the triples $\langle$\texttt{dbp:Kromsdorf}, \texttt{dbp:located\_in}, \texttt{dbp:Germany}$\rangle$ and $\langle$\texttt{lgd:240111203}, \texttt{lgd:is\_in}, \texttt{lgd:51477}$\rangle$ with the triples $\langle$\texttt{village}, \texttt{dbp:located\_in}, \texttt{country}$\rangle$ and $\langle$\texttt{village}, \texttt{lgd:is\_in}, \texttt{country}$\rangle$, respectively.
Using the predicate-proximity-graph, AutoAlign can learn the similarity between predicates from two KGs that represent the same relationships, e.g., the predicates \texttt{dbp:located\_in} and \texttt{lgd:is\_in}. 
\color{black}
Capturing predicate similarity in different KGs via a predicate-proximity-graph has a few \textbf{challenges}. First, each entity often has multiple types, which makes it difficult to directly align the predicates through the entity types. For example, the entity \texttt{Germany} may have multiple entity types \{\texttt{thing, place, location, country}\} in a KG. Second, different KGs may correspond to different sets of entity types, e.g., in another KG, the entity \texttt{Germany} may have entity types \{\texttt{place, country}\}. Hence, in the predicate-proximity-graph, the head entity and the tail entity may be replaced by multiple entity types. To address the above challenges, we propose two algorithms for aggregating multiple types of an entity and highlighting the most distinctive entity type (e.g., focusing more on \texttt{country} than on \texttt{thing}) via \emph{pseudo-type embedding}, which is a representation obtained from aggregating multiple entity types' information according to their importance.
Such an approximate predicate algorithm provides an automatic way of aligning predicates between two KGs, which not only complements the latent type information but also can be optimized by further joint learning for better predicate embeddings.

To achieve entity alignment, we exploit attribute triples and propose \emph{attribute character embeddings} for capturing the similarity between attributes; entities that have similar attributes should also be similar. Before our work, there was one study that has proposed an embedding for attributes \cite{sun2017cross}. However, it only uses the attribute types for computing embedding, which loses all the content information of the attributes and is ineffective in capturing attribute (dis)similarity. \textit{We are the first to propose attribute embedding that is based on the textual contents of the attributes \cite{trisedya2019entity}}.
Capturing the similarity of attributes of two KGs attribute similarity between entities in two KGs helps the attribute embedding to yield a unified embedding space for two KGs. This enables us to use attribute embeddings to shift the entity embeddings of two KGs into the same vector space and hence allows the entity embeddings to capture the similarity between entities from two KGs. 

\textbf{With the above two components, we achieve the first fully automatic method for KG alignment.} The contributions of this paper are as follows.

\begin{enumerate}[label=\textbf{C\arabic*}:] 
	\item We propose AutoAlign, a fully automatic KG alignment method that aligns two KGs with no seed alignments required (neither predicate nor entity seed alignments). Specifically, we propose automatic predicate alignment algorithms, automatic entity alignment algorithms, and a scheme to perform joint learning of entity, attribute, and predicate embeddings.
	\item We are the first to propose attribute embedding based on the textual contents of the attributes, which enables automatic entity alignment.
	\item We propose an automatic predicate alignment algorithm, enabled by two techniques: (i) we use a predicate-proximity-graph powered by large language models to capture predicates as relationships of entity types, and (ii) we use pseudo-type embeddings that aggregate multiple entity types in the proximity graph as the vector representation for predicates. 
	\item We conduct an extensive experimental study, which shows that our method is highly effective while being fully automatic. 
	Compared to existing methods, which all rely on manually crafted seeds, AutoAlign outperforms the best baseline by up to 10.65\% in hits@10.
\end{enumerate}
\color{black}

This paper is an extended version of our earlier conference paper~\cite{trisedya2019entity}. There, we presented the basic idea of the attribute character embeddings (\textbf{C2}). However, the method in the previous paper~\cite{trisedya2019entity} requires manually crafted predicate alignments. Specifically, it uses edit distance to compute the similarity score between predicates, and manual inspection is required to remove false positives. In this journal extension,
we have made substantial new contributions. First, we propose novel algorithms to align predicates without seed alignments including exploiting the most recent large language models (\textbf{C3}). Second, we propose a scheme to put all these components together and perform joint learning of entity, attribute and predicate embeddings, achieving a fully automatic KG alignment method (\textbf{C1}). Third, we have conducted a much more comprehensive experimental study, comparing with more baselines such as GNN-based ones, and using more recent benchmarks with more realistic and much larger datasets (\textbf{C4}).

\color{black}
\section{Related Work} 
	\label{kba-related-work}

This section discusses the related work, including knowledge graph embedding methods and knowledge graph alignment methods.

\subsection{Knowledge Graph Embedding Methods} 
\label{related-work-kbe-models}
Knowledge graph embedding methods are to address KB completion tasks \cite{socher2013reasoning,trisedya2021gcp} that aim to predict missing entities (i.e., head entity or tail entity) or relations (i.e., predicates) based on triples in a knowledge graph. 
Knowledge graphs have garnered attention in various research domains. \cite{yang2020graphdialog} integrates external knowledge bases and captures dialogue semantics in end-to-end task-oriented dialogue systems. \cite{distiawan2019neural,ding2021prototypical} focus on capturing the relations between entities in sentences using knowledge graphs. \cite{distiawan2018gtr,trisedya2020sentence} leverage knowledge graphs for data-to-text generation.
TransE \cite{bordes2013translating}, is a simple yet effective knowledge graph embedding method. TransE represents a relationship between a pair of entities as a \emph{translation} between the embeddings of the entities: a triple that consists of $\langle \texttt{head, predicate, tail} \rangle$ denoted as $\langle h, p, t \rangle$ should hold the property of $\mathbf{h + p \approx t}$. This representation indicates that the 
embedding vector of the tail entity approximately equals the embedding vector of the head entity after a vector sum operation of the embedding vector of predicate $\mathbf{p}$.
A scoring function $f(\mathbf{h,t}) = \left\lVert \mathbf{h+p-t}\right\rVert_2$ is used to measure the plausibility of a triple.
There are subsequent variants of TransE such as TransH \cite{wang2014knowledge} and TransR \cite{lin2015learning} by capturing more complex relationships.

Another popular paradigm of KG embedding is via graph neural networks~\cite{kipf2017semi}. Recently, graph embedding based on Transfomers~\cite{yun2019graph} have been proposed~\cite{hu2020heterogeneous,mei2022relation}. These methods learn entity embeddings via information propagation between nodes in a graph. 

\subsection{Knowledge Graph Alignment Methods} 
\label{related-work-kba-models}

The embedding methods above aim to preserve the structural information of the entities, i.e., entities that share similar neighbor structures in the KB should have similar representations in the embedding space. The advancement of such embedding methods motivates researchers to study embedding-based entity alignment. Chen et al. \cite{chen2017multilingual} propose MTransE, an embedding-based method for multilingual entity alignment based on TransE. 
In the follow-up work, Chen et al. \cite{chen2017multigraph} propose a generalized affine-map-based method to improve the alignment method of MTransE for handling various forms of invertible transformations.
Sun et al. \cite{sun2017cross} propose a joint attribute embedding method (called JAPE) for cross-lingual entity alignment. JAPE improves the alignment by capturing the correlations of attributes via the attribute type similarity. 

As the other popular paradigm for KG embedding, many GNN-based entity alignment methods have been proposed \cite{cao2019multi,sun2020knowledge,wu2019relation,ye2019a}. 
To improve the performance of GNN for entity alignment, existing methods \cite{cao2019multi} combine entity embeddings and attribute type embeddings in the computation of GNN. 
AttrGNN \cite{liu2020exploring} performs entity alignment by focusing on attributes, values, and structures. UPLR \cite{li2022uncertainty} effectively learns alignment information from pseudo-labeled datasets containing noisy data. 
MHGCN \cite{gao2021mhgcn} employs a multiview highway graph convolutional network to consider entity alignment from different views. 
These works share a common goal of leveraging different types of information present in knowledge graphs to improve entity alignment performance.
However, our proposed method stands out by being fully automated and leveraging both predicate and attribute triples without requiring seed alignments or multiple views.

In summary, all these entity alignment methods require manually crafted seed alignments. There has been no existing work on automatic predicate alignment. Due to the significant drawbacks of using seed alignments, our work focuses on automatic KG alignment methods.
A recent survey \cite{zhang2022benchmark} shows that translation-based methods outperform GNN-based ones in accuracy and efficiency, and our method AutoAlign is a translation-based one.

\blue{There are studies on cross-lingual entity alignment such as \cite{sun2017cross,mangrulkar2022multilingual,huang2022multilingual,ahuja2020language}, which find various applications in e-commerce.
Recent studies \cite{zhang2022benchmark} have shown that aligning KGs originating from different sources but the same language is another common setting with many real-life industrial applications \cite{zhang2020industry,liu2023cross}. In this paper, we focus on this latter setting, which conducts monolingual KG alignment. 
Moreover, \cite{zhang2022benchmark} has shown that translating two languages into one and then performing monolingual alignment achieves similar accuracy to what direct cross-lingual entity alignment gets.
Therefore, our method can also be applied to cross-lingual entity alignment via the translation approach.}
\color{black}

\section{Preliminary} 
	\label{kba-preliminary}
	
We start with the problem definition. A knowledge graph $\mathcal{G}$ consists of relation triples and attribute triples. A relationship triple is in the form of $\left<h,p,t\right>$, where $p$ is a relationship (\textit{predicate}) between two entities $h$~(\textit{head}) and $t$ (\textit{tail}). An attribute triple is in the form of $\left<h,p,v\right>$ where $v$ is an attribute value of entity $h$ with respect to the predicate $p$. 

Given two knowledge graphs $\mathcal{G}_1$ and $\mathcal{G}_2$, the task of entity alignment aims to find every pair $\langle h_1, h_2 \rangle$ where $h_1 \in \mathcal{G}_1$, $h_2 \in \mathcal{G}_2$, and $h_1$ and $h_2$ represent the same real-world entity. We use an embedding-based method that assigns a continuous representation for each element of a triple in the forms of $\mathbf{\left<h,p,t\right>}$ and $\mathbf{\left<h,p,v\right>}$, where the bold-face letters denote the vector representations of the corresponding element.

Our proposed method is built on top of a translation-based embedding method. We first discuss translation-based embedding methods and their limitations when being used for entity alignment before presenting our method.

\begin{figure*}[t!]
	\begin{center}
		\includegraphics[width=.8\textwidth, height=4.1in]{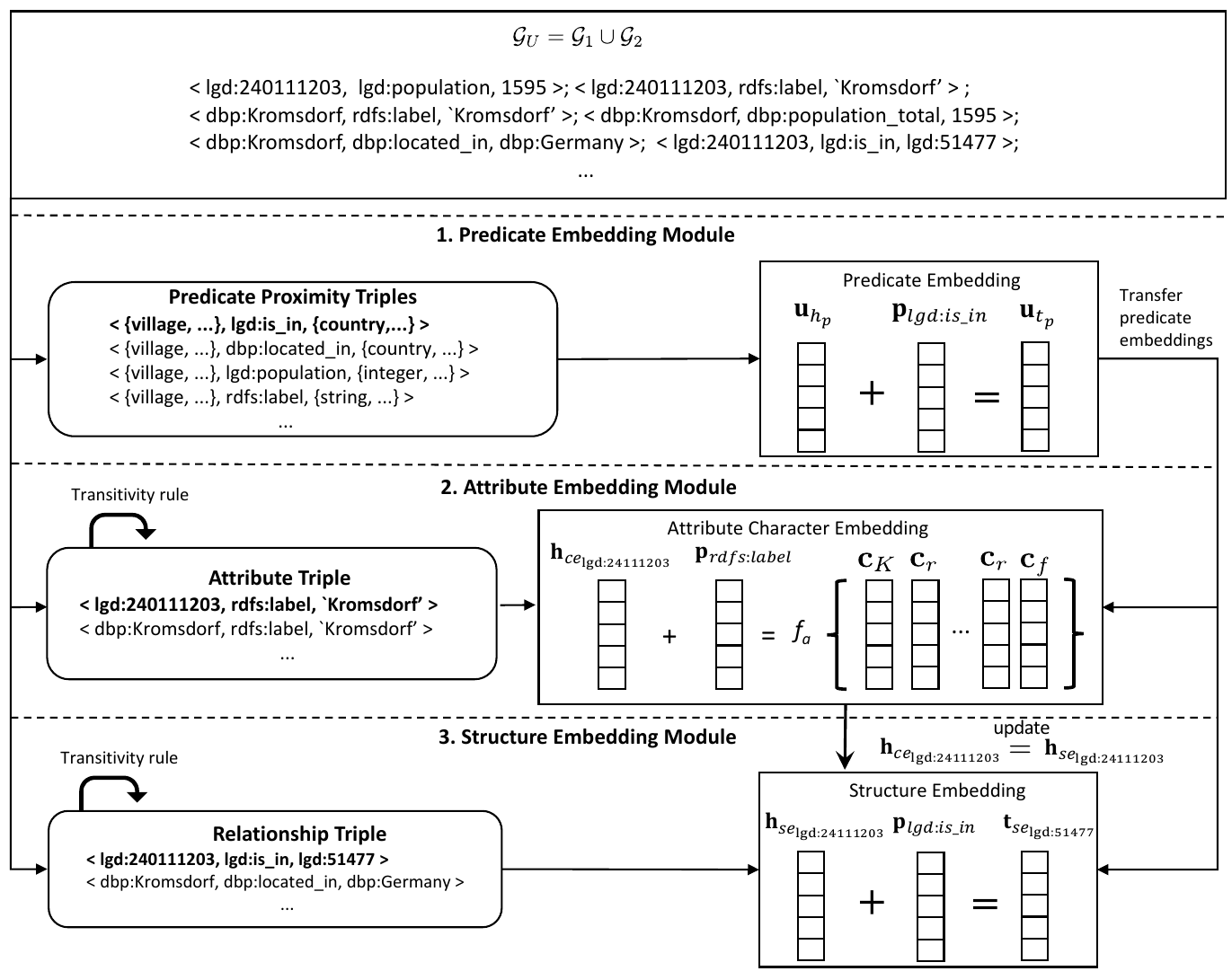}
	\end{center}
	\vspace{-3mm}
	\caption{\label{fig-kba-overall-module} Overview of our proposed AutoAlign method for entity alignment.}
\end{figure*}

\subsection{Translation-based Embedding Method} 
	\label{kba-transe}

Given a relationship triple $\langle h,p,t \rangle$, a translation-based embedding method, such as TransE~\cite{bordes2013translating}, suggests that the embedding of the tail entity $t$ should be close to the embedding of the head entity $h$ plus the embedding of the relationship $p$, i.e., $\mathbf{h + p \approx t}$. Such an embedding method aims to preserve the structural information of the entities, i.e., entities that share similar neighbor structures in a knowledge graph should have similar representations in the embedding space.
We refer to the modeling of the structural information as \emph{structure embedding} and the modeling should preserve the translation property of $\mathbf{h + p \approx t}$.
To learn the structure embedding, TransE minimizes a margin-based objective function $\mathcal{J}_{SE}$:
\begin{align}
	\mathcal{J}_{SE}&=\sum_{t_r\in \mathcal{T}_r} \sum_{t_r^\prime \in \mathcal{T}_r^\prime} \max\left(0,\left[ \gamma+  f(t_r)- f(t_r^\prime) \right] \right) \label{eq-kba-ori-jse}\\
	f(t_r)&=\left\Vert \mathbf{h+p-t} \right\Vert_2 \\
	\mathcal{T}_r&=\{\langle h, p, t \rangle | \langle h,p,t \rangle \in \mathcal{G}\} \\
	{\mathcal{T}_r}^\prime&=\left\{\left<h^\prime,p,t\right>\left.\right|h^\prime\in \mathcal{E}\right\} \cup\left\{\left< h,p,t^\prime\right>\left.\right|t^\prime\in \mathcal{E}\right\}
\end{align}
Here, $\left\Vert \textbf{x} \right\Vert_2$ is the L2-Norm of vector $\textbf{x}$,  $\gamma$ is a margin hyperparameter, $\mathcal{T}_r$ is the set of valid relation triples, and $\mathcal{T}_r^\prime$ is the set of corrupted relation triples ($\mathcal{E}$ is the set of entities in $\mathcal{G}$). The corrupted triples are used as negative samples created by replacing the head or tail entity of a valid triple in $\mathcal{T}_r$ with a random entity.

\begin{figure*}[t!]
	\centering
	\includegraphics[width=.8\textwidth,height=0.9in]{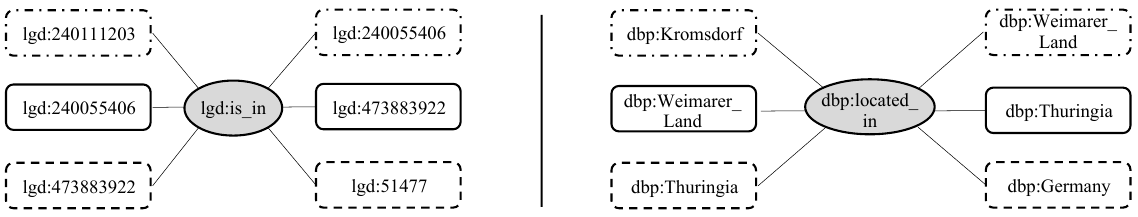}
	\caption{Predicate graph of the similar relationship \texttt{lgd:is\_in} (left) and \texttt{dbp:located\_in} (right) in two KGs. Each type of dotted line represents the similar entity types in two predicate graphs. }
	\label{fig-kba-predicate-graph-raw}
	\vspace{-3mm}
\end{figure*}

The advantages of structure embeddings drive further studies of embedding-based entity alignment. However, a straightforward implementation of structure embedding for entity alignment has limitations: the entity embeddings computed on different KGs may fall in different spaces, where similarity cannot be computed directly. Existing methods~\cite{chen2017multilingual,sun2017cross,zhu2017iterative} address this limitation by computing a transition matrix to map the embedding spaces of different KGs into the same space, as discussed earlier. However, such methods require manually collecting a seed set of aligned entities from the different KGs to compute the transition matrix, which does not scale and is vulnerable to the quality of the manually crafted seed aligned entities.

Next, we detail our method to address these limitations.

\color{black}
\section{Proposed Method} 
	\label{kba-proposed-model}
We give an overview of AutoAlign in Section \ref{kba-solution-overview}. We then explain the various components of AutoAlign, which include the \emph{predicate embedding} module in Section \ref{kba-predicate-embedding}, the \emph{structure embedding} module in Section \ref{kba-structure-embedding}, the \emph{attribute embedding} module in Section \ref{kba-attribute-character-embedding}, the joint learning scheme in Section \ref{kba-joint-learning}, entity alignment in Section \ref{kba-entity-alignment}, and triple enrichment in Section \ref{kba-transitivity-rule}. We discuss the scalability of AutoAlign in Section \ref{sec:Scalability}.

\subsection{Overview of AutoAlign}
\label{kba-solution-overview}
AutoAlign has three core embedding modules: the predicate embedding, the attribute embedding, and the structure embedding. Fig.~\ref{fig-kba-overall-module} gives an overview and shows the interaction between the three modules.

In order to execute the embedding-based KG alignment, it's crucial to ensure that both predicate and entity embeddings of two KGs coexist in the identical vector space. To meet this criterion, we start by simply making a union of the two knowledge graphs, denoted as $\mathcal{G}_U = \mathcal{G}_1 \cup \mathcal{G}_2$. This put all the triples from both KGs together in their original form. 
Note that $\mathcal{G}_U$ is different from $\mathcal{G}_M$ as the entities in $\mathcal{G}_U$ are not aligned yet. 
We will obtain three types of triples from $\mathcal{G}_U$: predicate-proximity triples $\mathcal{T}_p$, relation triples $\mathcal{T}_r$, and attribute triples $\mathcal{T}_a$.
In the set of predicate-proximity triples, each predicate corresponds to a connection between entity types. 
For instance, the set of predicate-proximity triples may contain triples such as $\langle$\texttt{village}, \texttt{dbp:located\_in}, \texttt{country}$\rangle$ and $\langle$\texttt{village}, \texttt{lgd:is\_in}, \texttt{country}$\rangle$. 
These triples help train a predicate embedding module (elaborated in Section \ref{kba-predicate-embedding}) to capture the similarity between predicates from two KGs, such as \texttt{dbp:located\_in} and \texttt{lgd:is\_in}. 
This process not only generates a unified embedding space for predicates but also captures their similarity between the two KGs. Predicate embeddings are then utilized for computing attribute and structure embeddings.

The structure and entity embeddings are obtained from the set of relation triples $\mathcal{T}_r$ as discussed in Section \ref{kba-structure-embedding}, and the attribute embedding, as discussed in Section \ref{kba-attribute-character-embedding}, leveraging the set of attribute triples $\mathcal{T}_a$. Initially, due to the unique naming schemes in each KG, entity embeddings from $\mathcal{G}_1$ and $\mathcal{G}_2$ exist in two different vector spaces. However, we unify the attribute embeddings derived from attribute triples $\mathcal{T}_a$ into a common vector space. This unification is via the character embeddings learned from attribute strings, which can manifest similarity despite their origin from different KGs. The obtained attribute embeddings are then utilized to align the entity embeddings into a common vector space, thus enabling the entity embeddings to reflect the similarity between entities across both KGs.

Upon acquiring the embeddings for all entities in $\mathcal{G}_1$ and $\mathcal{G}_2$, the entity alignment module (explained in Section \ref{kba-entity-alignment}) identifies every pair $\left<h_1, h_2\right>$, with $h_1 \in \mathcal{G}_1$ and $h_2 \in \mathcal{G}_2$, that has a similarity score exceeding a threshold $\beta$.

To bolster the effectiveness of AutoAlign, we implement the transitivity rule to expand an entity's properties, thereby fostering a more resilient attribute embedding for gauging entity similarities. This procedure is discussed in Section \ref{kba-transitivity-rule}.

\subsection{Predicate Embedding Module}
\label{kba-predicate-embedding}

The same predicates from two KGs typically connect the same entity type, albeit in different surface forms. Consider the example in Fig.~\ref{fig-kba-predicate-graph-raw}, where the predicate \texttt{lgd:is\_in} in LinkedGeoData and the predicate \texttt{dbp:located\_in} in DBpedia connect three entity pairs. 
In LinkedGeoData, the predicate \texttt{lgd:is\_in} connects $\langle$\texttt{lgd:240111203,lgd:240055406}$\rangle$, $\langle$\texttt{lgd:240055406, lgd:473883922}$\rangle$, and $\langle$\texttt{lgd:473883922,lgd:51477}$\rangle$. 
Meanwhile, in DBpedia, the predicate \texttt{dbp:located\_in} connects $\langle$\texttt{dbp:Kromsdorf,dbp:Weimarer\_Land}$\rangle$, $\langle$\texttt{dbp:Weimarer\_Land,dbp:Thuringia}$\rangle$, and $\langle$\texttt{dbp:-Thuringia,dbp:Germany}$\rangle$. 
Here, the entity pairs from both knowledge graphs correspond to the same real-world entity pairs. For instance, the head and tail entities of the entity pair $\langle$\texttt{lgd:240111203,lgd:240055406}$\rangle$ and $\langle$\texttt{dbp:Kromsdorf,dbp:Weimarer\_Land}$\rangle$ represent a village named Kromsdorf and a district named Weimarer Land, respectively. 
However, due to the distinct naming schemes of the two knowledge graphs, entity embedding methods may not capture this similarity. Applying an entity embedding method to the raw knowledge graph could result in the predicate embeddings of \texttt{lgd:is\_in} and \texttt{dbp:located\_in} being placed in different vector spaces.

In our previous work \cite{trisedya2019entity}, a semi-automatic predicate alignment process was employed to tackle this challenge. This involved renaming akin predicates from the two KGs via a uniform naming scheme, thereby generating a shared vector space for relationship embeddings. To pinpoint similar predicates across the two KGs, we leveraged string edit distance and subsequently examed any false positives manually. However, such a method presents limitations in real-world applications, given its reliance on manual intervention for aligning predicates, or creating `seed alignments', across the two KGs.

To address the above problem, AutoAlign introduces a fully automatic predicate alignment procedure by learning predicate embeddings from a \emph{predicate-proximity-graph} of two KGs. 
The predicate-proximity-graph replaces entities with their corresponding entity types. Through learning the graph, we can effectively identify predicate similarities between two KGs without the need to manually compare predicates' surface forms.  
The automatic predicate alignment is detailed in Sections \ref{sec:ppgc} and \ref{sec:ppg_learning}.

\subsubsection{Predicate-proximity-graph Construction}
\label{sec:ppgc}

A predicate-proximity-graph is essentially a graph depicting the relationships between \emph{entity types} rather than \emph{entities}. 
Entity types indicate the broad categories of entities, which automatically link different entities.
Even if some predicates have different surface forms (e.g., \texttt{lgd:is\_in} v.s. \texttt{dbp:located\_in}), we can effectively identify them as being similar by learning the predicate-proximity-graph. This is because the head/tail entities of these predicates usually have similar entity types (e.g., $\langle$\texttt{place, lgd:is\_in,country}$\rangle$, $\langle$\texttt{place, dbp:located\_in,country}$\rangle$).
To create the predicate-proximity-graph,
we start with $\mathcal{G}_U$ the union of the two KGs' triples in their original forms, and then replace each triple's head and tail entities with their respective entity types. Next, the entity types are obtained through two steps: \emph{entity type extraction} and \emph{type alignment enabled by large language models}, which are described as follows.

\noindent
\textbf{1) Entity Type Extraction.} 
We extract entity types by taking the values of the \texttt{rdfs:type} predicate for every entity from each KG. It's common for each entity to have multiple types. For instance, the entity \texttt{Germany} might have several entity types like {\texttt{place, location, country}} within a KG. Furthermore, varying KGs may adopt different schemes of entity types, e.g., in another KG, the entity \texttt{Germany} might have entity types like {\texttt{place, country}}. To accommodate these variations, within the predicate-proximity-graph, we substitute both the head and tail entities of each triple in a knowledge graph with a collection of entity types. For example, we replace the triples $\langle$\texttt{dbp:Kromsdorf}, \texttt{dbp:located\_in}, \texttt{dbp:Germany}$\rangle$ with the triples $\langle\mathcal{U}_{kromsdorf}$, \texttt{dbp:located\_in}, $\mathcal{U}_{germany}\rangle$. Here, $\mathcal{U}{x}$ is a set of types for entity $x$, e.g., $\mathcal{U}{germany}$= \{\texttt{thing, place, location, country}\}\footnote{We provide the details of automatically obtain and align the entity types from two KGs in Appendix \ref{appx:obtain_type}}.

\noindent
\textbf{2) Type Alignment enabled by Large Language Models.}
After obtaining the entity types of the two KGs, we need to perform an alignment between the entity types of two KGs. This is because the types of two KGs may refer to the same meaning but using different surface forms, e.g., person v.s. people. Therefore, we need to align such types into one for the best effect of predicate-proximity-graph training. In previous work, we manually align the types since the number of the types is not large. With the recent breakthrough of large language models such as ChatGPT and Claude \cite{floridi2020gpt,scao2022bloom,thoppilan2022lamda}, we can eliminate such manual effort and make it fully automatic.

Specifically, we use Claude\footnote{https://www.anthropic.com/index/introducing-claude}, which is a free and powerful large language model. We construct a prompt as input to Claude as follows:

\begin{center}
\fbox{\parbox{0.95\columnwidth}{
\textit{
``Now you are an expert in linguistics and knowledge graphs. I will give you two sets of words, indicating the entity types from two knwoledge graphs. You need to identify all the word pairs from the two sets that are synonyms. For example, if the first set has the word `people' and the second set has the word `person', you need to identify the two words being synonyms and return me the pair (people, person).
Now the following are the two sets:
Set 1: \{people, music,...\}
Set 2: \{person, thing,...\}
Please return all the pairs that are synonyms from the two sets reagarding entity types. Do not output the pairs if they are exactly the same. Remember you only need to return the pairs, each pair in one line. Each pair contain two types, one from Set 1 and another from Set 2, in the format (type1, type2).''}}}
\end{center}

The type set in \emph{Set 1} and \emph{Set 2} can be filled with the types extracted from corresponding KGs, and the other contents are fixed for any two KGs. We feed the prompt to Claude, and it will return the pairs of types that are similar to each other in the format of (type1, type2). 
After identifying type1 of $\mathcal{G}_1$ is similar to type2 in $\mathcal{G}_2$, we replace all the type2 by type1 so that similar types are represented by the same surface form, e.g., replace ``people'' with ``person'' so that two KGs both use the type ``person''. This way we obtain type alignment without human intervention. 

After extracting the entity types for each KG and aligning the extracted types enabled by LLMs, we obtain the predicate-proximity-graph that has the same number of triples as $\mathcal{G}_U$, with each triple's entity being replaced by their types. 


\subsubsection{Module Learning}
\label{sec:ppg_learning}

To capture the predicate similarity, the module should focus on the most distinctive entity types, e.g., emphasizing \texttt{country} more than \texttt{thing}. We propose two ways for aggregating multiple entity types: 1) weighted sum function, and 2) attention-based function. 
In the experimental study, we show that the attention-based function works better.


\noindent
1) \textbf{Weighted Sum Function:} Given the entity type embedding $\mathbf{U}_x=(\vec{\mathbf{z}}_0, \vec{\mathbf{z}}_1, \cdots, \vec{\mathbf{z}}_M)$ of entity type $z \in \mathcal{U}_x$ with $M$ types, we calculate the \emph{pseudo-type embedding} $\mathbf{u}$ as follows.

\begin{equation}
\label{eq-weight-u}
\mathbf{u}=\sum_{i=0}^{M} w_i \vec{\mathbf{z}}_i,~w_i = \text{softmax}\left(\frac{l_i}{a_i r_i}{k_i}\right)
\end{equation}

Here, the weight $w_i$ controls the distribution of $\vec{\mathbf{z}}_i$, which is the vector representation of an entity type $z$ in $\mathcal{U}_x$. To give a larger weight to the most distinctive type, we use $l_i$, which is the level of specificity of an entity type in WordNet \cite{fellbaum1998wordnet}. An entity type with a deeper level of specificity has a larger value of $l_i$, e.g., \texttt{country} has a deeper level of specificity than \texttt{thing}. This way, the predicate embedding module can emphasize the most distinct type. We normalize the weight using three variables. The first is $a_i$, which is the number of attributes in $\mathcal{U}_x$. The second is $r_i$, which is the number of occurrences of type $z$ in a KG. Intuitively, an entity type that appears in almost all entities, such as \texttt{thing}, is less distinctive, and the predicate embedding module should filter this entity type to obtain a better predicate representation. The third variable is $k_i$, which is the number of KGs that contain the entity type $z$ to indicate the agreements between two KGs on the entity type $z$. Lastly, we use $\text{softmax}$ to transform the weights into probability distributions for each type in $\mathcal{U}_x$.

\noindent
2) \textbf{Attention-based Function:} In contrast to the most distinctive entity types, there may be some ``noise'' entity types that contribute little when learning the meaning of a predicate. For example, in the set of entity types for entity \texttt{Germany}, $\mathcal{U}_{germany}$= {\texttt{thing, place, location, country}}, the entity type \texttt{thing} $\in \mathcal{U}_{germany}$ may be less relevant to the triple $\langle$\texttt{dbp:Kromsdorf}, \texttt{dbp:located\_in}, \texttt{dbp:Germany}$\rangle$. In this case, \texttt{thing} can be seen as a "noise" entity type that is not essential for learning the representation of predicate \texttt{dbp:located\_in}.

To better capture the importance of different entity types for a predicate, we propose an attention-based algorithm, which allows the module to adaptively evaluate the entity type weight and ignore potential type noise. We calculate the attention weight $z_i$ of the $i$-th entity type as:
\begin{equation}
z_i=softmax(U^T_x W_z \vec{\mathbf{z}}_i)
\end{equation}
where $W_z$ denotes the trainable weight matrix of entity type embedding $ \vec{\mathbf{z}}_i$. The final pseudo-type embedding $\mathbf{u}$ is obtained through a weighted sum of all corresponding entity type vectors $ \vec{\mathbf{z}}_i$ representing different semantic meanings:
\begin{equation}
\label{eq-attention-u}
\mathbf{u}=\sum_{i=0}^{M}z_i  \vec{\mathbf{z}}_i
\end{equation}
where $ \vec{\mathbf{z}}_i$ is the embedding of the $i$-th entity type.

The pseudo-type embeddings computed by Equations \ref{eq-weight-u} or \ref{eq-attention-u} are used as the proximity entity embeddings, which are used next to train the predicate embeddings as follows. 

We use the pseudo-type embeddings $\mathbf{u_{h_{p}}}$ and $\mathbf{u_{t_{p}}}$ to represent the corresponding head entity and tail entity in the predicate proximity triples $\mathcal{T}_p$, respectively. We then compute the predicate embeddings by minimizing the following objective function:
\begin{eqnarray}
\mathcal{J}_{PE}=\sum_{t_p\in \mathcal{T}_p} \sum_{t_p^\prime \in \mathcal{T}_p^\prime} \max\left(0,\left[ \gamma+  f(t_p)- f(t_p^\prime) \right] \right)\\
f(t_p)=\left\Vert \mathbf{u_{h_{p}}+p-u_{t_{p}}} \right\Vert_2
\label{eq-J-PE}
\end{eqnarray}
where $t_p$ is a triple in the predicate-proximity-graph and $t_p^\prime$ is a corrupted triple (i.e., for negative samples) generated based on the predicate-proximity-graph. Here, $\mathbf{u_{h_{p}}}$ and $\mathbf{u_{t_{p}}}$ can be obtained using the above two functions, Eq. \ref{eq-weight-u} and Eq. \ref{eq-attention-u}.
We use AutoAlign-W and AutoAlign-A to denote AutoAlign utilizing Eq. \ref{eq-weight-u} and Eq. \ref{eq-attention-u}, respectively.

Note that the procedure discussed can be extended to compute embeddings for attribute predicates by modifying Eq. \ref{eq-J-PE} to replace the entity types of the tail entity in relation triples with the literal types (e.g., string, integer, and long data type) of attribute values in attribute triples.
Through optimizing the objective function, AutoAlign cultivates a unified predicate embedding space from two knowledge graphs. This method empowers us to transition these embeddings into the learning of structure and attribute embeddings.

\subsection{Structure Embedding Module}
\label{kba-structure-embedding}

Our method of computing the structure embeddings is built based on TransE. Although TransE typically assigns equivalent weights to each neighbor when computing an entity's embeddings, we adjust TransE to give different weights to an entity's different neighbors. The underlying reasoning is to assign higher weights to neighbors that are linked by predicates already aligned, as they serve as a significant indicator for entity alignment.

As illustrated in Table \ref{table-kba-example}, we can categorize the predicates of KGs into three groups. The first group comprises the already aligned predicates, such as \texttt{geo:lat, geo:long,} and \texttt{rdfs:label}, which adhere to the predicate naming scheme convention\footnote{https://www.w3.org/TR/rdf-schema/} in knowledge graphs. The second group contains implicitly aligned predicates, such as \texttt{lgd:is\_in} and \texttt{dbp:located\_in}. These predicates are beneficial for entity alignment if we can identify the alignment between them. We address this issue with our predicate embedding module (see Section~\ref{kba-predicate-embedding}). The final group consists of non-aligned predicates, such as \texttt{lgd:alderman} and \texttt{dbp:district}. These predicates do not aid entity alignment and are treated as noise.

To mitigate the influence of noise, we adjust TransE by incorporating a weight factor $\alpha$ to govern the learning of embeddings across the triples. As a result, the entity embedding approach can sift out non-aligned triples grounded on non-aligned predicates. To deduce the structure embedding in AutoAlign, we aim to minimize the objective function $\mathcal{J}_{SE}$, which is adapted from Eq.~\eqref{eq-kba-ori-jse}, as follows:
%
\begin{align}
\mathcal{J}_{SE}&=\sum_{t_r\in\ \mathcal{T}_r} \sum_{t_r^\prime \in\ \mathcal{T}_r^\prime} \max\left(0,\gamma+ \alpha \left( f(t_r)- f(t_r^\prime) \right) \right)\label{eq-kba-jse}\\
\alpha&=\frac{count(r)}{\left| \mathcal{T} \right|}
\end{align}
where $\mathcal{T}_r$ represents the set of valid relation triples, $\mathcal{T}r^\prime$ denotes the set of corrupted relation triples, $count(r)$ is the occurrence count of relationship $r$, and $\left| \mathcal{T} \right|$ is the total number of triples in the merged KG $\mathcal{G}{U}$. Typically, the occurrence count of already aligned and implicitly aligned predicates is greater than that of non-aligned predicates (as aligned predicates are present in both KGs, while non-aligned predicates only appear in one of the KGs). Therefore, our weighting algorithm enables the embedding method to learn more effectively from the aligned triples. For instance, in the triples shown in Table \ref{table-kba-example}, the weight $\alpha$ assists the embedding method in prioritizing relationships like \texttt{rdfs:label}, \texttt{geo:lat}, and \texttt{geo:long} ($\alpha = 2/12$ for each of these predicates) over relationships such as \texttt{lgd:alderman} or \texttt{dbp:district} ($\alpha = 1/12$ for each of these predicates).

\subsection{Attribute Embedding Module} 
\label{kba-attribute-character-embedding}

For attribute embedding, we construe the attribute predicate $p$ as a transition from the head entity $h$ to the attribute value $v$. An attribute might be represented in multiple forms across two KGs. For instance, \texttt{50.9989} versus \texttt{50.9988888889} as an entity's latitude, or \texttt{"Barack Obama"} against \texttt{"Barack Hussein Obama"} as a person's name. We adopt a compositional function to code the attribute value and establish the relationship of each component in an attribute triple as $\mathbf{h}~+~\mathbf{p}~\approx~f_a(v)$. Here, $f_a(v)$ signifies a compositional function, and $v$ denotes a sequence of the characters of the attribute value $v~=~\left\{ c_1,c_2,c_3,..., c_t \right\}$. This compositional function compiles the attribute value into a single vector, thereby associating similar attribute values to a like vector representation. We present three compositional functions as described below.

\textbf{Sum Compositional Function} (\textbf{SUM}): The first compositional function is defined as the sum of all character embeddings of the attribute value.
\begin{equation}
f_a(v) = \mathbf{c_1 + c_2 + c_3 + ... + c_t}
\end{equation}
where $\mathbf{c_1, c_2, ..., c_t}$ represent the character embeddings of the attribute value. While this compositional function is straightforward, it suffers from a major limitation: two strings with the same character set but arranged in a different order will have the same vector representation (i.e., order invariant). For instance, two coordinates, \texttt{"50.15"} and \texttt{"15.05"}, will result in the same vector representation.

\textbf{LSTM-based Compositional Function} (\textbf{LSTM}). To address the above problem, we propose an LSTM-based compositional function. This function uses LSTM networks to encode a sequence of characters into a single vector. We use the final hidden state of the LSTM networks as a vector representation of the attribute value.
\begin{equation}
f_a(v) = f_{lstm}({\mathbf{c_1, c_2, c_3, ..., c_t}})
\end{equation}
where $f_{lstm}$ is an LSTM network \cite{hochreiter1997long}.

\textbf{N-gram-based Compositional Function} (\textbf{N-gram}). LSTM-based compositional function handles the order invariant problem. However, it only considers the unigram features of a string. To capture rich compositional information of a string, we further propose an N-gram-based compositional function as an alternative to the above two compositional functions. Here, we use the summation of the n-gram combination of the attribute value.
\begin{equation}
f_a(v)=\sum_{n=1}^{N}\left(\frac{\sum_{i=1}^{l}\sum_{j=i}^{n}\mathbf{c_j}}{t-i-1}\right)
\end{equation}
where $N$ indicates the maximum value of $n$ used in the n-gram combinations ($N = 10$ in our experiments), and $l$ is the length of the attribute value.

To learn the attribute embedding, we minimize the following objective function $\mathcal{J}_{CE}$:

\begin{equation}
\begin{split}
\mathcal{J}_{CE}&=\sum_{t_a\in \mathcal{T}_a} \sum_{t_a^\prime \in \mathcal{T}_a^\prime} \max\left(0,\left[ \gamma+  \alpha \left( f(t_a)- f(t_a^\prime) \right) \right] \right)\label{eq-kba-jce}\\
f(t_a)&=\left\Vert \mathbf{h}+\mathbf{p}-f_a(v) \right\Vert_2,~\mathcal{T}_a=\{\langle h,p,v \rangle \in \mathcal{G}_U\}\\
{\mathcal{T}_a}^\prime&=\left\{\left.\left<h^\prime,p,v\right>\right|h^\prime\in \mathcal{E}_U\right\}\cup\left\{\left<h,p,v^\prime\right>|v^\prime\in \mathcal{A}_U\right\}
\end{split}
\end{equation}

Here, $\mathcal{T}_a$ is the set of valid attribute triples from the training dataset, and $\mathcal{T}_a^\prime$ is the set of corrupted attribute triples ($\mathcal{A}_U$ is the set of attributes in $\mathcal{G}_U$). The corrupted triples are used as negative samples by replacing the head entity with a random entity or the attribute with a random attribute value. $f(t_a)$ is the plausibility score computed based on the embedding of the head entity $h$, the embedding of the attribute predicate $p$, and the vector representation of the attribute value computed using function $f_a(v)$.

\subsection{Joint Learning of the Embeddings} 
\label{kba-joint-learning}

AutoAlign jointly learns the predicate embeddings, the structure embeddings, and the attribute embeddings. The proposed method first trains over the predicate-proximity-graph to yield the unified predicate embedding space. 
AutoAlign then uses these predicate embeddings to jointly learn the structure and attribute embeddings.
However, the attribute embedding module yields a unified embedding space for two knowledge graphs but lacks structure information. On the other hand, the structure embedding module may yield different embedding space for two knowledge graphs. Thus, we use the attribute embedding $\mathbf{h_{ce}}$ to shift the structure embedding $\mathbf{h_{se}}$ into the same vector space by minimizing the following objective function $\mathcal{J}_{SIM}$:
\begin{equation}
\mathcal{J}_{SIM}=\sum_{s \in \mathcal{G}_1 \cup \mathcal{G}_2} \left[1- \cos(\mathbf{h_{se}},\mathbf{h_{ce}}) \right]
\end{equation}
Here, $\cos(\mathbf{h_{se}},\mathbf{h_{ce}})$ is the cosine similarity of vector $\mathbf{h_{se}}$ and $\mathbf{h_{ce}}$. As a result, the structure embedding captures the similarity of entities between two KGs based on entity relationships, while the attribute embedding captures the similarity of entities based on attribute values. The overall objective function of the joint learning is:
\begin{equation}
\mathcal{J} = \mathcal{J}_{PE} + \mathcal{J}_{SE} + \mathcal{J}_{CE} + \mathcal{J}_{SIM}
\end{equation}

\subsection{Entity Alignment} 
\label{kba-entity-alignment}

The existing embedding-based entity alignment methods are supervised when obtain the resulting embeddings since they need seed alignments to learn entity alignments from two knowledge graphs. Unlike the existing methods, AutoAlign captures the similarity between entities from two knowledge graphs by learning a unified entity embedding space via predicate and attribute embeddings. AutoAlign does not need seed alignments. Our joint learning embedding scheme lets similar entities from $\mathcal{G}_1$ and $\mathcal{G}_2$ have close vector representations. Thus, the resultant embeddings can be used for entity alignment. We compute the following equation for entity alignment.
\begin{equation}
h_{map}=\argmax_{h_2 \in \mathcal{G}_2} \cos(\mathbf{h}_1, \mathbf{h}_2) \label{eq-kba-hmap}
\end{equation}
Given an entity $h_1 \in \mathcal{G}_1$, we compute the similarity between $h_1$ and all entities $h_2 \in \mathcal{G}_2$; $\left< h_1, h_{map} \right>$ is the expected pair of aligned entities. We use a similarity threshold $\beta$ to filter the pairs of entities that are too dissimilar to be aligned.

\subsection{Triple Enrichment via Transitivity Rule}
\label{kba-transitivity-rule}

In translation-based embedding methods such as TransE, the embedding of an entity is learned by aggregating information from its immediate neighbors (i.e., one-hop neighbors). These methods may implicitly learn the multi-hop relationships between entities via information propagation after many training epochs. However, the information propagation of the multi-hop relationship is weak. On the other hand, the explicit inclusion of multi-hop relationships (e.g., transitive relationships) increases the number of attributes and related entities for each entity, which helps identify the similarity between entities. For example, given triples $\langle$\texttt{dbp:Emporium\_Tower, :locatedIn, dbp:London}$\rangle$ and $\langle$\texttt{dbp:London, :country, dbp:England}$\rangle$, we can infer that \texttt{dbp:Emporium\_Tower} has a relationship (i.e., \texttt{":locatedInCountry"}) with \texttt{dbp:England}. In fact, this information can be used to enrich the related entity \texttt{dbp:Emporium\_Tower}. We treat the one-hop transitive relation as follows. Given transitive triples $\left< h_1,p_1, t_1 \right>$ and $\left< t_1, p_2, t_2 \right>$, we interpret $p_1.p_2$ as a relation from head entity $h_1$ to tail entity $t_2$. Therefore, the relationship between these transitive triples is defined as $\mathbf{h_1 + (p_1 . p_2) \approx t_2}$. The objective functions of the transitivity-enhanced embedding methods are adapted from the Eq. \eqref{eq-kba-jse} and Eq. \eqref{eq-kba-jce} by replacing the relationship vector $\mathbf{p}$ with $\mathbf{p_1.p_2}$.

\subsection{Scalability Discussion}
\label{sec:Scalability}
AutoAlign has three main modules, predicate embedding module, structure embedding module and attribute embedding module; their most time-consuming operations are to iterate through the corresponding training samples (i.e., the triples) on the proximity graph using Equation \ref{eq-J-PE}, the relation graph using Equation \ref{eq-kba-jse}, and the attribute graph using Equation \ref{eq-kba-jce}, respectively. The numbers of triples of the three graphs are all upper-bounded by the total number of edges (i.e., triples) in the two KGs, which we denote as $\mathcal{M}$. Therefore, the time complexities of the predicate embedding module, structure embedding module and attribute embedding module are all $\mathcal{O(M)}$. The triple enrichment via transitivity rule modifies part of the triples without bringing new ones, and does not increase the complexity. Therefore, the time complexity of AutoAlign is still $\mathcal{O(M)}$. 

As analysed by \cite{zhang2022benchmark}, translation-based methods \cite{chen2017multilingual,bordes2013translating,IPTransE2017} typically have the same time complexity, $\mathcal{O(M)}$, since their most time-consuming operations are to iterate through all the training samples (i.e., triples) in the two KGs. GNN-based models also have $\mathcal{O(M)}$ time complexity and require loading the whole graph into memory due to the message passing mechanism. Therefore, AutoAlign has the same time complexity as state-of-the-art KG alignment methods. 

We have conducted experiments to compare the running time of AutoAlign to two recent baselines AttrGNN \cite{liu2020exploring} and UPLR \cite{li2022uncertainty}. We observe that the running time of AutoAlign is twice that of AttrGNN and half that of UPLR. This is acceptable and reasonable, which is consistent with our complexity analysis above.   

\color{black}

\section{Experiments} 
\label{kba-experiments}

We evaluate AutoAlign from three different aspects. First, we show the performance of AutoAlign in entity alignment, which is the main task in this paper. Second, we show that our predicate embedding module effectively aligns predicates from different knowledge graphs. Third, we show that the resulting embeddings of AutoAlign preserve the structure information of knowledge graphs, enabling them to be used in broader applications such as KG completion.
\vspace{-3mm}

\subsection{Datasets}
\label{kba-experiments-dataset}

\begin{table}[t!]
	\centering
	\caption{Statistics of the datasets for entity alignment.}
	\label{table-dataset_statistics}
	\resizebox{0.99\columnwidth}{!}{%
        \color{black}
		\begin{tabular}{l|c|c|c|c|c}
			\toprule[2pt]
			\midrule
			\multicolumn{1}{c|}{Subset} & \multicolumn{1}{p{3.57em}|}{Unique\newline{}entities} & \multicolumn{1}{l|}{Predicates} & \multicolumn{1}{p{5.715em}|}{Relationship\newline{}triples} & \multicolumn{1}{p{4.215em}|}{Attribute\newline{}triples} & \multicolumn{1} {p{3.215em}}{Entity\newline{}types}\\
			\midrule
			\midrule
			\multicolumn{6}{c}{\textbf{DW-NB}}\\
			\midrule
			DBpedia & 84,911 &  545   & 203,502 & 221,591 & 93\\
			Wikidata & 86,116 & 703   & 198,797 & 223,232 & 257\\
			\midrule
			\midrule
			\multicolumn{6}{c}{\textbf{DY-NB}} \\
			\midrule
			DBpedia & 58,858 &  211   & 87,676 & 173,520 & 50\\
			Yago  & 60,228 & 91    & 66,546 & 186,328  & 61\\
			\midrule
			\bottomrule[2pt]
		\end{tabular}%
	}
	\vspace*{-3mm}
\end{table}%

We evaluate our method on the latest comprehensive benchmark for KG alignment, \emph{DWY-NB} \cite{zhang2022benchmark}, which consists of two datasets \emph{DW-NB} and \emph{DY-NB}.
The two KGs of DW-NB are subsets of DBpedia~\cite{auer2007dbpedia} and Wikidata~\cite{vrandecick2014wikidata}, respectively. The two KGs of DY-NB are subsets of DBpedia~\cite{auer2007dbpedia} and Yago~\cite{hoffart2013yago2}, respectively.  
Specifically, DW-NB has more than 84,911 unique entities and contains 50,000 aligned entities, DY-NB has more than 58,858 unique entities and contains 15,000 aligned entities.
$36\%$ of the aligned entities have different entity names, which makes the datasets more realistic and the entity alignment task more challenging. 
To compare with baselines that require entity seeds. We randomly select 50\% of the seed alignment as a test set, and the rest of them are used as entity seeds.
The statistics of the datasets are summarized in Table \ref{table-dataset_statistics}. 

\subsection{Implementation Details}
\label{kba-experiments-hyperparameters}

We use grid search to find the best hyperparameters for AutoAlign. We choose the embeddings dimensionality $d$ among $\{50, 75, 100, 200\}$, the learning rate of the Adam optimizer among $\{0.001, 0.01, 0.1\}$, and the margin $\gamma$ among $\{1, 5, 10\}$. We train AutoAlign with a batch size of $100$ and a maximum of $400$ epochs.
We compare with representative state-of-the-art methods, and have used the hyper-parameters suggested by their corresponding papers.

\begin{table*}[htp]
	\centering
	\caption{The effect of the amount of seed entity alignments on EA performance in terms of Hits@k (\%). The numbers with bold/underlined indicate the highest/sub-optimal values in each group compared to baseline methods.}
	\label{table-exp-1-results}
	\resizebox{0.90\textwidth}{!}{%
        \color{black}
		\begin{tabular}{c|l|rr|rr|rr|rr|rr|rr}
			\toprule[2pt]
			\midrule
			\multicolumn{2}{c|}{\multirow{2}[4]{*}{Method}} & \multicolumn{2}{c|}{Seed: 0\%} &  \multicolumn{2}{c|}{Seed: 10\%} & \multicolumn{2}{c|}{Seed: 20\%} & \multicolumn{2}{c|}{Seed: 30\%} & \multicolumn{2}{c|}{Seed: 40\%} & \multicolumn{2}{c}{Seed: 50\%} \\
		  \cmidrule{3-14}  
            \multicolumn{2}{c|}{} &\multicolumn{1}{l}{Hits@1} & \multicolumn{1}{c|}{Hits@10} & \multicolumn{1}{l}{Hits@1} & \multicolumn{1}{l|}{Hits@10} & \multicolumn{1}{l}{Hits@1} & \multicolumn{1}{l|}{Hits@10} & \multicolumn{1}{l}{Hits@1} & \multicolumn{1}{l|}{Hits@10} & \multicolumn{1}{l}{Hits@1} & \multicolumn{1}{l|}{Hits@10} & \multicolumn{1}{l}{Hits@1} & \multicolumn{1}{l}{Hits@10} \\
			\midrule
			\midrule
			\multicolumn{14}{c}{DW-NB} \\
			\midrule
			\multirow{8}[-2]{*}{\begin{sideways}Translation-based\end{sideways}} & MTransE & N/A & N/A & 2.82  & 10.45 & 5.42  & 18.72 & 7.88  & 25.75 & 10.42 & 31.44 & 12.98 & 36.00 \\
			& IPTransE & N/A & N/A & 5.98  & 13.45 & 7.54  & 18.78 & 12.90 & 24.61 & 16.32 & 32.86 & 23.54 & 35.98 \\
			& BootEA & N/A & N/A & 8.12  & 16.15 & 12.54 & 20.13 & 17.92 & 28.38 & 21.46 & 35.16 & 25.44 & 37.57 \\
			& TransEdge & N/A & N/A & 22.98 & 48.12 & 38.29 & 56.22 & 45.27 & 68.95 & 49.26 & 75.25 & 54.85 & 79.68 \\
			& \underline{JAPE} & N/A & N/A & 4.62  & 7.87  & 8.62  & 14.43 & 12.57 & 19.96 & 17.20 & 27.32 & 19.91 & 30.63 \\
			& \underline{MultiKE} & N/A & N/A & 80.25 & 87.58 & 82.56 & 88.92 & 84.06 & 90.05 & 84.87 & 91.24 & 85.21 & 95.06 \\
			& \underline{AttrE} & \underline{87.98} & 95.80 & \underline{87.98} & 95.80 & \underline{87.98} & 95.80 & \underline{87.98} & 95.80 & \underline{87.98} & 95.80 & 87.98 & 95.80 \\
			& \underline{AutoAlign-W} & 87.81 & \underline{95.86} & 87.81 & \underline{95.86} & 87.81 & \underline{95.86} & 87.81 & \underline{95.86} & 87.81 & \underline{95.86} & 87.81 & \underline{95.86} \\
			& \underline{AutoAlign-A} & \textbf{88.73} & \textbf{96.91} & \textbf{88.73} & \textbf{96.91} & \textbf{88.73} & \textbf{96.91} & \textbf{88.73} & \textbf{96.91} & \textbf{88.73} & \textbf{96.91} & \textbf{88.73} & \textbf{96.91} \\
			\midrule
			\multirow{11}[2]{*}{\begin{sideways}GNN-based\end{sideways}} & MuGNN & N/A & N/A & 13.49 & 37.79 & 20.96 & 49.28 & 26.92 & 56.77 & 31.09 & 61.43 & 34.41 & 64.96 \\
			& AliNet & N/A & N/A & 14.58 & 31.46 & 18.55 & 35.84 & 24.34 & 50.46 & 28.39 & 55.46 & 35.31 & 58.22 \\
			& KECG  & N/A & N/A & 18.95 & 34.17 & 24.32 & 40.78 & 30.24 & 48.66 & 35.29 & 52.12 & 39.40 & 62.31 \\
			& \underline{GCN-Align} & N/A & N/A & 12.40 & 30.18 & 20.04 & 41.56 & 24.76 & 48.52 & 29.02 & 53.43 & 31.80 & 56.20 \\
			& \underline{HGCN} & N/A & N/A & 58.08 & 62.15 & 63.14 & 68.15 & 78.97 & 86.51 & 84.25 & 90.75 & 88.54 & 91.54 \\
			& \underline{GMNN} & N/A & N/A & 71.32 & 74.24 & 75.34 & 79.23 & 80.98 & 82.23 & 82.67 & 85.87 & 84.59 & 88.64 \\
			& \underline{RDGCN} & N/A & N/A & 59.22 & 62.98 & 64.22 & 68.98 & 79.02 & 87.12 & 85.34 & 90.45 & 88.21 & 93.23 \\
			& \underline{CEA}  & N/A & N/A & 50.13 & 52.31 & 63.25 & 64.12 & 80.32 & 84.21 & 84.34 & 85.54 & 86.58 & 88.34 \\
			& \underline{MRAEA} & N/A & N/A & 53.75 & 54.74 & 64.58 & 66.12 & 81.54 & 85.97 & 83.54 & 86.02 & 84.06 & 87.55 \\
			& \underline{NMN}  & N/A & N/A & 51.45 & 59.78 & 68.21 & 72.54 & 84.03 & 88.21 & 85.65 & 90.54 & \underline{88.69} & 95.46 \\
                & AttrGNN & N/A & N/A & 45.79  & 78.28 &  51.67 & 82.85 & 54.65 & 84.30 & 59.48 & 86.18 & 62.08 & 88.74 \\
                & UPLR &  0.34 & 1.62 &  0.34 & 1.62 &  0.34 & 1.62 &  0.34 & 1.62 &  0.34 & 1.62  &  0.34 & 1.62 \\
			\midrule
			\midrule
			\multicolumn{14}{c}{DY-NB} \\
			\midrule
			\multirow{8}[-2]{*}{\begin{sideways}Translation-based\end{sideways}} & MTransE & N/A & N/A & 0.01  & 0.15  & 0.01  & 0.39  & 0.08  & 0.68  & 0.08  & 1.39  & 0.13  & 1.89 \\
			& IPTransE & N/A & N/A & 1.54  & 9.87  & 5.67  & 25.76 & 14.55 & 36.45 & 15.77 & 45.81 & 17.33 & 52.18 \\
			& BootEA & N/A & N/A & 2.15  & 14.19 & 8.47  & 38.15 & 15.77 & 48.32 & 17.22 & 57.15 & 19.24 & 58.14 \\
			& TransEdge & N/A & N/A & 22.98 & 47.50 & 37.85 & 64.85 & 48.98 & 72.15 & 58.95 & 76.54 & 62.49 & 78.54 \\
			& \underline{JAPE} & N/A & N/A & 0.70  & 1.83  & 1.57  & 3.37  & 1.40  & 3.27  & 1.37  & 1.77  & 2.37  & 4.97 \\
			& \underline{MultiKE} & N/A & N/A & 81.87 & 88.05 & 82.11 & 89.26 & 84.97 & 90.84 & 87.22 & 92.05 & 89.25 & 93.58 \\
			& \underline{AttrE} & \underline{90.44} & 94.23 & \underline{90.44} & 94.23 & \underline{90.44} & 94.23 & \underline{90.44} & 94.23 & \underline{90.44} & 94.23 & 90.44 & 94.23 \\
			& \underline{AutoAlign-W} & 90.42 & \underline{94.35} & 90.42 & \underline{94.35} & 90.42 & \underline{94.35} & 90.42 & \underline{94.35} & 90.42 & \underline{94.35} & 90.42 & 94.35 \\
			& \underline{AutoAlign-A} & \textbf{91.27} & \textbf{95.62} & \textbf{91.27} & \textbf{95.62} & \textbf{91.27} & \textbf{95.62} & \textbf{91.27} & \textbf{95.62} & \textbf{91.27} & \textbf{95.62} & \textbf{91.27} & \textbf{95.62} \\
			\midrule
			\multirow{11}[2]{*}{\begin{sideways}GNN-based\end{sideways}} & MuGNN & N/A & N/A & 19.16 & 51.41 & 27.40 & 62.69 & 31.60 & 68.56 & 34.73 & 71.24 & 37.15 & 74.07 \\
			& AliNet & N/A & N/A & 13.54 & 28.53 & 14.25 & 31.69 & 25.39 & 58.31 & 28.98 & 56.12 & 34.59 & 64.12 \\
			& KECG  & N/A & N/A & 11.19 & 23.65 & 14.89 & 27.25 & 20.95 & 34.48 & 22.81 & 35.44 & 24.71 & 37.15 \\
			& \underline{GCN-Align} & N/A & N/A & 8.56  & 25.09 & 17.88 & 43.88 & 24.36 & 53.43 & 31.29 & 62.44 & 33.56 & 67.88 \\
			& \underline{HGCN} & N/A & N/A & 52.54 & 64.51 & 65.87 & 77.40 & 71.14 & 85.64 & 71.45 & 85.64 & 74.54 & 87.48 \\
			& \underline{GMNN} & N/A & N/A & 62.34 & 70.34 & 64.32 & 67.34 & 75.57 & 77.47 & 78.65 & 82.65 & 82.34 & 85.62 \\
			& \underline{RDGCN} & N/A & N/A & 53.13 & 65.30 & 67.28 & 78.21 & 74.54 & 85.22 & 77.45 & 87.43 & 78.67 & 89.45 \\
			& \underline{CEA}  & N/A & N/A & 55.24 & 58.97 & 64.35 & 65.42 & 74.56 & 78.42 & 77.78 & 80.95 & 78.91 & 83.24 \\
			& \underline{MRAEA}& N/A & N/A  & 52.46 & 53.20 & 60.33 & 64.54 & 73.71 & 78.52 & 74.25 & 78.66 & 76.22 & 80.15 \\
			& \underline{NMN} & N/A & N/A & 55.74 & 64.78 & 62.54 & 70.54 & 75.87 & 80.54 & 84.55 & 88.69 & \underline{90.78} & \underline{94.77} \\
                & AttrGNN  & N/A & N/A & 77.21 & 88.03 & 79.44 & 89.76 & 80.16 & 90.19 & 81.31 & 90.84 & 83.98 & 91.95 \\
                & UPLR & 89.84  & 93.11 & 89.84  & 93.11 & 89.84  & 93.11 & 89.84 & 93.11 & 89.84 & 93.11 & 89.84 & 93.11 \\
			\midrule
			\midrule[2pt]
			\multicolumn{14}{l}{* Methods that use attribute triples are \ul{underlined}. The rest tables and figures follow this convention.}\\
			\multicolumn{14}{l}{* AttrE, AutoAlign-W and AutoAlign-A do not use any seed alignments.}
		\end{tabular}%
	}
	\vspace{-2mm}
\end{table*}%

\subsection{Compared Methods} 
\label{kba-experiments-models}
AutoAlign has two ways for aggregating
multiple entity types, weighted sum function and attention-based function as described in Section \ref{sec:ppg_learning}. We use \textbf{AutoAlign-W} and \textbf{AutoAlign-A} to represent AutoAlign with weighted sum function and attention-based function respectively.
Other compared existing entity alignment methods are described below.

	 \textbf{MTransE}~\cite{chen2017multilingual} is the state-of-the-art embedding-based alignment method built on top of TransE. MTransE learns a transition matrix from seed alignments to yield a unified embedding space from two KGs.
	 \textbf{IPTransE}~\cite{IPTransE2017} is an improved version of TransE. IPTransE adopts two soft strategies to add newly-aligned entities to the seeds to mitigate error propagation. 
	 \textbf{BootEA}~\cite{bordes2013translating} models EA as a one-to-one classification problem where the counterpart of an entity is regarded as the label of the entity. It iteratively learns the classifier via bootstrapping from both labeled and unlabeled data. 
	 \textbf{TransEdge}~\cite{sun2019transedge} proposes an edge-centric translational embedding method addressing the deficiency of TransE in that its relation predicate embeddings are entity-independent. 
	 \textbf{JAPE}~\cite{sun2017cross} is another state-of-the-art embedding-based entity alignment method built on top of TransE. It combines the relation triples with masked attribute triples. A masked attribute triple is an attribute triple in which its object is replaced by its data type.
	 \textbf{MultiKE}~\cite{MultiKE2019} uses multi-view learning on various kinds of features. The embedding module of MultiKE divides the features of KGs into three subset views. Entity embeddings are learned for each view and then combined.
	 \textbf{AttrE}~\cite{trisedya2019entity} is the first method that makes use of attribute values and the only EA method that needs no seed alignments.
	 \textbf{MuGNN} \cite{cao2019multi} is the state-of-the-art embedding-based entity alignment  built on top of GCN, which uses two GCN as different channels to encode a KG. One channel is for completing missing links in a KG, and the other channel is for filtering unnecessary entities.
	 \textbf{AliNet}~\cite{sun2020knowledge} learns entity embeddings by a controlled aggregation of entity neighborhood information, and shares similar neighborhood structures by considering both direct and distant neighbors. 
	 \textbf{KECG}~\cite{KECG2019} aims to reconcile the issue of structural heterogeneity between KGs by jointly training both a GAT-based cross-graph module and a TransE-based knowledge embedding module.
	 \textbf{GCN-Align}~\cite{GCN-Align2018} is the first study on GNN-based EA, which learns entity embeddings from structural information of entities and exploits attribute triples by treating them as relation triples.
	 \textbf{HGCN}~\cite{HGCN2019} explicitly utilizes relation representation to improve the alignment process in EA. It incorporates the relation information by jointly learning entity and relation predicate embeddings.
	 \textbf{GMNN}~\cite{GMNN2019} formulates the EA task as graph matching between two topic entity graphs. It uses a graph matching module to model the similarity of two topic entity graphs, which indicates the probability of the two corresponding entities being aligned.
     \textbf{RDGCN}~\cite{RDGCN2019} utilizes relation information and extends GCNs with highway gates to capture the neighborhood structural information. It differs from HGCN in that it incorporates relation information by the attentive interaction.
     \textbf{CEA}~\cite{CEA2020} proposes a collective EA method which considers the dependency of alignment decisions among entities. It uses structural, semantic, and string signals to capture different aspects of the similarity between entities in the source and the target KGs, which are represented by three separate similarity matrices.
    \textbf{MRAEA}~\cite{MRAEA2020} considers meta relation semantics including relation predicates, relation direction, and inverse relation predicates, in addition to structural information learned from merely the structure of relation triples.
    \textbf{NMN}~\cite{NMN2020} aims to tackle the structural heterogeneity between KGs. The method learns both the KG structure information and the neighborhood difference so that the similarities between entities can be better captured.
    \textbf{AttrGNN}~\cite{liu2020exploring} performs entity alignment by combining attribute graph learning, value graph learning, and structure graph learning, and selects the best performance by comparing different combinations.
    \textbf{UPLR}~\cite{li2022uncertainty} constructs pseudo-labeled datasets containing noisy data and leverage the graph attention nets to capture the similarities between two KGs.
\vspace{-3mm}

\subsection{Entity Alignment Results} 
\label{kba-entity-alignment-results}

This experiment evaluates the performance of EA while varying the amount of seed entity alignments used for training from $10\%$ to $50\%$ of the total available set of seed entity alignments (50,000 for DW-NB and 7,500 for DY-NB).
We fix the test set in all the settings for all the models. That is, the baseline methods with different ratios of seed alignments will have the same test set. This setting ensures that the results in different settings are comparable and hence valid for evaluating the model performance.
We evaluate the performance of AutoAlign (note that it does not need any seed alignments) using $\mathbf{Hits@k} (k=1,10)$ (i.e., the proportion of correctly aligned entities ranked in the top $k$ predictions). A higher value indicates better performance. 

Table~\ref{table-exp-1-results} shows the results on the DWY-NB benchmark datasets \cite{zhang2022benchmark}. Some of the results of the compared methods are from \cite{zhang2022benchmark}. 
Note that since our methods do not require entity seeds, the results of our model (and UPLR, which also does not require entity seeds) are the same for different seed ratios.
We observe that the two variations of AutoAlign, AutoAlign-W and AutoAlign-A, are significantly better than all the other methods. 
The performance of AutoAlign-A is better than AutoAlign-W, which shows the importance to capture both the distinctive and noisy entity types, as done by AutoAlign-A.
The underlined methods from both translation- and GNN-based methods exploit attribute triples. The methods that exploit attribute triples achieve much better performance than the methods that do not. 

When no seed is provided (Seed: $0\%$), the baselines that require entity seeds simply cannot run, while our methods can still get the great performance.
When seeds are available, other methods can run but perform considerably worse than our methods. For example, AutoAlign-A outperforms the best performing baseline, MultiKE, by 10.65\% in hits@10 in the DW-NB dataset (96.91 v.s. 87.58).
The performance of these methods get better with more seed alignments available, but they are still considerably worse than our methods, especially when fewer seed alignments are available.


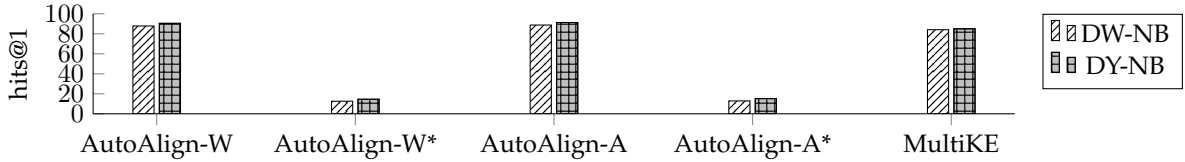
\begin{figure*}[htp]
    \centering
	\begin{tikzpicture}
	\begin{groupplot}[
	width = 0.76\textwidth,
	height = 0.16\textwidth,
	legend pos= outer north east 
	]
	
	\nextgroupplot[
	enlarge x limits=0.08,
	axis x line*=bottom,
	axis y line*=left,
	ybar,
	ylabel=hits@1,
	ymin=0,
	ymax=100,
	xtick={AutoAlign-W, AutoAlign-W*,  AutoAlign-A, AutoAlign-A*,   MultiKE}, 
	symbolic x coords={{AutoAlign-W}, {AutoAlign-W*},  {AutoAlign-A}, {AutoAlign-A*},  {MultiKE}},
	bar width=8pt,
	]
	\addplot[pattern=north east lines] coordinates { (AutoAlign-W, 87.81) (AutoAlign-W*, 12.58) (AutoAlign-A, 88.73) (AutoAlign-A*, 12.88)  (MultiKE, 84.06) };
	\addplot[color=black, fill=black!25, postaction={pattern=grid}] coordinates {(AutoAlign-W, 90.42) (AutoAlign-W*, 14.55) (AutoAlign-A, 91.27) (AutoAlign-A*, 15.21)  (MultiKE, 84.97) };
	
	\addlegendimage{pattern=north east lines}
	\addlegendentry{ DW-NB }
	\addlegendimage{color=black, fill=black!25, postaction={pattern=grid}}
	\addlegendentry{ DY-NB }
	
	\end{groupplot}
	\end{tikzpicture}
	\vspace{-2mm}
	\caption{The effect of attribute embedding module.}
	\label{fig-exp-2-results}
\end{figure*}
\vspace{-3mm}

\begin{figure*}
\centering
	\begin{tikzpicture}
	\begin{groupplot}[
	group style = {group size = 2 by 1, horizontal sep=1.5cm},
	width = 0.46\textwidth,
	height = 0.16\textwidth,
	]
	
	
	\nextgroupplot[
	enlarge x limits=0.3,
	axis x line*=bottom,
	axis y line*=left,
	ybar,
	ylabel=hits@1,
	ymin=75,
	ymax=100,
	xlabel=DW-NB,
	xtick={SUM, LSTM, N-gram},
	symbolic x coords={{SUM},{LSTM},{N-gram}},
	bar width=15pt,
	nodes near coords,
	every node near coord/.append style={rotate=90, anchor=west},
	]
	\addplot[color=red, fill=red!25] coordinates {(SUM,78.68) (LSTM,86.73) (N-gram,87.98)};
	\addplot[color=blue, fill=blue!25] coordinates {(SUM,77.54) (LSTM,86.79) (N-gram,87.81)};
	\addplot[color=orange, fill=orange!25] coordinates {(SUM,80.02) (LSTM,87.53) (N-gram,88.73)};
	
	\nextgroupplot[
	enlarge x limits=0.3,
	axis x line*=bottom,
	axis y line*=left,
	ybar,
	ylabel=hits@1,
	ymin=75,
	ymax=100,
	xlabel=DY-NB,
	xtick={SUM, LSTM, N-gram},
	symbolic x coords={{SUM},{LSTM},{N-gram}},
	bar width=15pt,
	nodes near coords,
	every node near coord/.append style={rotate=90, anchor=west},
	legend style={at={($(3,3)+(2cm,2cm)$)},legend columns=1,fill=none,draw=none,anchor=center,align=left},
	legend to name=lg
	]
	\coordinate (c2) at (rel axis cs:1,1);
	\addplot[color=red, fill=red!25] coordinates {(SUM,83.17) (LSTM,87.41) (N-gram,90.44)};
	\addplot[color=blue, fill=blue!25] coordinates {(SUM,82.5) (LSTM,87.32) (N-gram,90.42)};
	\addplot[color=orange, fill=orange!25] coordinates {(SUM,84.63) (LSTM,88.78) (N-gram,91.27)};
	\addlegendimage{red!25}
	\addlegendentry{Semi-automatic predicate alignment (AttrE)}
	\addlegendimage{blue!25}
	\addlegendentry{Proximity-graph-based predicate alignment (AutoAlign-W)}
	\addlegendimage{blue!25}
	\addlegendentry{Proximity-graph-based predicate alignment (AutoAlign-A)}
	\end{groupplot}
	\node[below] at (current bounding box.south){\pgfplotslegendfromname{lg}};
	\end{tikzpicture}
	\vspace{-2mm}
	\caption{The effect of predicate embedding module.}
	\label{fig-kba-predicate-alignment}
\end{figure*}
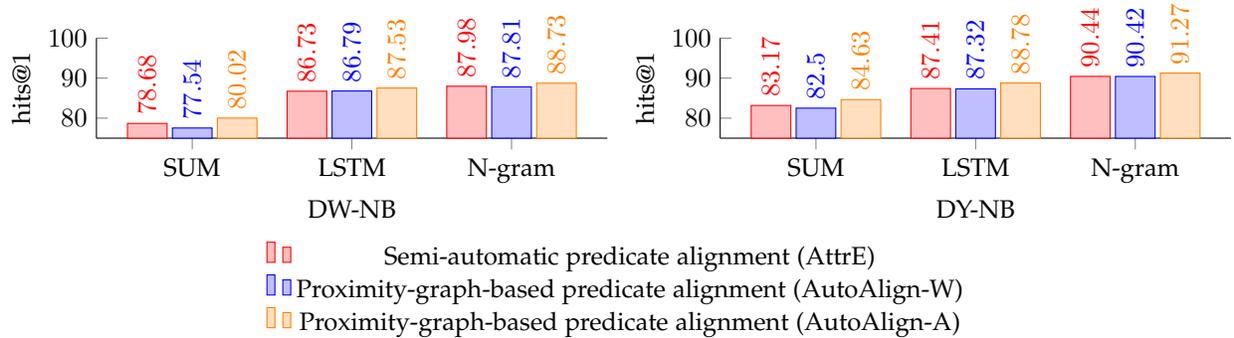
\vspace{-0.3cm}

\subsection{Ablation Study}
We conduct ablation tests from two perspectives to evaluate AutoAlign: the effect of attribute embedding module and the effect of predicate embedding module.

\subsubsection{Effect of Attribute Embedding Module} 
\label{kba-discussion-attribute}
To evaluate the effect of using attribute triples, we create a version of AutoAlign-W that does not use attribute triples to compute the entity embeddings, i.e., it only uses relation triples; we call this version AutoAlign-W*. Similarly, we create a version of AutoAlign-A that does not use attribute triples, which we call AutoAlign-A*. Figure~\ref{fig-exp-2-results} shows the performance of the four versions of AutoAlign on the benchmark measured by Hit@1. We can see that the performance of AutoAlign-W and AutoAlign-A are much higher than that of AutoAlign-W* and AutoAlign-A*, respectively. This shows that our idea of using attribute triples 
is highly effective. We also put the performance of MultiKE in the figure for comparison since MultiKE is the most accurate one among other existing methods; the proportion of seed entity alignments used for MultiKE is $30\%$. AutoAlign-W and AutoAlign-A both outperform MultiKE.


We also show the effect of different attribute embedding algorithms in Fig.~\ref{fig-kba-predicate-alignment}. Here SUM, LSTM, and N-gram denote three algorithms with different attribute embedding functions, as described in Section \ref{kba-attribute-character-embedding}. 
We see that the N-gram compositional function gives the best performance. This is because the N-gram compositional function better preserves string similarity when mapping attribute strings to their vector representations than the other two functions. 

\subsubsection{Effect of Predicate Embedding Module} 
\label{kba-discussion-predicate}

To evaluate the effect of predicate embedding module proposed in Section~\ref{kba-predicate-embedding}, we compare with the semi-automatic predicate alignment module in our previously proposed method AttrE \cite{trisedya2019entity}. 
From Fig.~\ref{fig-kba-predicate-alignment}, we see that the predicate embedding module helps our entity alignment method achieve comparable performance in terms of $hits@1$. 

The same predicate may be stored in different surface forms in the KGs, e.g., one KG has the attribute predicate \texttt{birth\_date} while the other KG has the attribute predicate \texttt{date\_of\_birth}. 
Previous methods exploit seed attribute predicate alignments and seed attribute alignments to address this difference. 
In comparison, our AutoAlign-W and AutoAlign-A do not need manual intervention, and they both yield  competitive results. In particular, AutoAlign-A achieves the best performance since it enriches the related entity types information via the attention mechanism. 
\vspace{-2mm}

%
%
\begin{table}[hbt!]
	\begin{center}
		\caption{The effect on  the accuracy of downstream link prediction task in terms of Hits@10 (\%). The numbers with bold/underline indicate the highest/sub-optimal values in each group compared to baseline methods.}
		\label{table-exp-4-results}
		\resizebox{0.9\columnwidth}{!}{%
    		\begin{tabular}{@{}lllllll}
                \toprule[2pt]
                \midrule
                \multicolumn{1}{@{}c|}{\multirow{2}[4]{*}{Method}} & \multicolumn{3}{c|}{DW-NB (seed)} & \multicolumn{3}{c}{DY-NB (seed)} \\
            \cmidrule{2-7}    \multicolumn{1}{c|}{} & \multicolumn{1}{c}{(10\%)} & \multicolumn{1}{c}{(30\%)} & \multicolumn{1}{c|}{(50\%)} & \multicolumn{1}{c}{(10\%)} & \multicolumn{1}{c}{(30\%)} & \multicolumn{1}{c}{(50\%)} \\
                \midrule[1.5pt]
                 \multicolumn{1}{l|}{\ul{AutoAlign-A}} & \multicolumn{1}{r}{\textbf{88.93}} & \multicolumn{1}{r}{\underline{88.93}} & \multicolumn{1}{r|}{\underline{88.93}} & \multicolumn{1}{r}{\textbf{98.82}} & \multicolumn{1}{r}{\underline{98.82}} & \multicolumn{1}{r}{\textbf{98.82}} \\
                \multicolumn{1}{l|}{\ul{MultiKE}} & \multicolumn{1}{r}{\underline{88.76}} & \multicolumn{1}{r}{\textbf{88.98}} & \multicolumn{1}{r|}{\textbf{89.52}} & \multicolumn{1}{r}{98.62} & \multicolumn{1}{r}{\textbf{98.87}} & \multicolumn{1}{r}{98.07} \\
                \multicolumn{1}{l|}{\ul{AttrE}} & \multicolumn{1}{r}{88.50} & \multicolumn{1}{r}{88.50} & \multicolumn{1}{r|}{88.50} & \multicolumn{1}{r}{\underline{98.75}} & \multicolumn{1}{r}{98.75} & \multicolumn{1}{r}{\underline{98.75}} \\
                \multicolumn{1}{l|}{\ul{AutoAlign-W}} & \multicolumn{1}{r}{88.41} & \multicolumn{1}{r}{88.41} & \multicolumn{1}{r|}{88.41} & \multicolumn{1}{r}{98.66} & \multicolumn{1}{r}{98.66} & \multicolumn{1}{r}{98.66} \\
                \multicolumn{1}{l|}{TransE} & \multicolumn{1}{r}{87.45} & \multicolumn{1}{r}{87.45} & \multicolumn{1}{r|}{87.45} & \multicolumn{1}{r}{98.42} & \multicolumn{1}{r}{98.42} & \multicolumn{1}{r}{98.42} \\
                \multicolumn{1}{l|}{TransEdge} & \multicolumn{1}{r}{85.27} & \multicolumn{1}{r}{85.71} & \multicolumn{1}{r|}{86.40} & \multicolumn{1}{r}{93.24} & \multicolumn{1}{r}{93.54} & \multicolumn{1}{r}{93.76} \\
                \multicolumn{1}{l|}{\ul{JAPE}} & \multicolumn{1}{r}{83.24} & \multicolumn{1}{r}{83.71} & \multicolumn{1}{r|}{83.09} & \multicolumn{1}{r}{75.03} & \multicolumn{1}{r}{75.32} & \multicolumn{1}{r}{75.66} \\
                \multicolumn{1}{l|}{IPTransE} & \multicolumn{1}{r}{81.06} & \multicolumn{1}{r}{81.23} & \multicolumn{1}{r|}{81.78} & \multicolumn{1}{r}{93.10} & \multicolumn{1}{r}{93.55} & \multicolumn{1}{r}{93.91} \\
                \multicolumn{1}{l|}{BootEA} & \multicolumn{1}{r}{80.41} & \multicolumn{1}{r}{80.90} & \multicolumn{1}{r|}{81.66} & \multicolumn{1}{r}{94.11} & \multicolumn{1}{r}{94.54} & \multicolumn{1}{r}{94.85} \\
                \multicolumn{1}{l|}{MTransE} & \multicolumn{1}{r}{80.10} & \multicolumn{1}{r}{80.33} & \multicolumn{1}{r|}{80.69} & \multicolumn{1}{r}{93.81} & \multicolumn{1}{r}{94.31} & \multicolumn{1}{r}{94.74} \\
                \midrule
                \midrule[2pt]
                \\
            \end{tabular}%
		}
	\end{center}
	\vspace{-5mm}
\end{table}
\subsection{Effect of the Alignment Method on KG embeddings} 
\label{kba-discussion2}

We evaluate the effect of AutoAlign on KG embeddings. 
This section experiments on how the quality of the KG embeddings obtained from EA methods is affected compared to the KG embeddings from pure KG embedding methods (TransE for translation-based and GCN for GNN-based methods) via downstream applications of KGs. 
Following previous studies in EA methods \cite{sun2020benchmarking}, we conduct experiments using a common downstream task \emph{link prediction} for this purpose, detailed as follows. 
The link prediction task aims to predict $t$ given $h$ and $r$ of a relation triple. 
Specifically, first, a relation triple is corrupted by replacing its tail entity with all the entities in the dataset. Then, the corrupted triples are ranked in ascending order by the plausibility score computed as $\boldsymbol{h} + \boldsymbol{r} - \boldsymbol{t}$. Since true triples (i.e., the triples in a KG) are expected to have smaller plausibility scores and rank higher in the list than the corrupted ones, hits@10 (whether the true triples are in the top-10) is used as the measure for the link prediction task.

Table~\ref{table-exp-4-results} shows the performance of link prediction on DW-NB and DY-NB with $10\%$, $30\%$, and $50\%$ of seed entity alignments. The performance increases with the amount of seed alignments but not significantly.
As mentioned earlier, the KG embeddings obtained from the KG alignment methods may not be optimized for downstream tasks. However, AutoAlign-A still achieves high performance, always in top-2 and top-1 in half of the cases, which shows that the learned predicate embeddings can also project entities into a unified embedding space. 

\vspace{-3mm}
\section{Conclusion and Future Work} 
\label{kba-conclusion}

We presented AutoAlign -- the first fully automatic method for KG alignment \red{enabled by large language models}. We proposed attribute character embeddings and predicate-proximity-graph embeddings powered by large language models to compute a unified vector space for the entity and predicate embeddings from two KGs. Experimental results show that AutoAlign outperforms the competitors consistently.
Further results on knowledge graph completion show that our joint learning of the entity, predicate, and attribute embeddings can capture the similarity between entities and predicates both within a KG and across KGs. 

AutoAlign demonstrates the potential of leveraging large language models to improve the performance of KG alignment (e.g., requiring less manual work, and incorporating the knowledge stored in large language models). In future work, we may investigate broader research domains based on graphs or hypergraphs \cite{jiang2018hyperx} that can benefit from large language models enabled KG alignment. For example, leveraging large language models to align KGs with domain-specific graphs (e.g., feature graphs in recommender systems \cite{su2021detecting,su2022detecting}, region graphs in point of interests learning \cite{zhao2023learning}) to enrich their representation ability. 


\section*{Acknowledgments}
This work is supported in part by NSFC Grant No. U1936205 and by grant from the Research Grants Council of the Hong Kong Special Administrative Region, China (No. CUHK 14217622).

\ifCLASSOPTIONcaptionsoff
  \newpage
\fi
\vspace{-3mm}
\bibliographystyle{IEEEtran}
\bibliography{bib/Trisedya_TKDE_AAAI19}

\begin{thebibliography}{10}
\providecommand{\url}[1]{#1}
\csname url@samestyle\endcsname
\providecommand{\newblock}{\relax}
\providecommand{\bibinfo}[2]{#2}
\providecommand{\BIBentrySTDinterwordspacing}{\spaceskip=0pt\relax}
\providecommand{\BIBentryALTinterwordstretchfactor}{4}
\providecommand{\BIBentryALTinterwordspacing}{\spaceskip=\fontdimen2\font plus
\BIBentryALTinterwordstretchfactor\fontdimen3\font minus
  \fontdimen4\font\relax}
\providecommand{\BIBforeignlanguage}[2]{{%
\expandafter\ifx\csname l@#1\endcsname\relax
\typeout{** WARNING: IEEEtran.bst: No hyphenation pattern has been}%
\typeout{** loaded for the language `#1'. Using the pattern for}%
\typeout{** the default language instead.}%
\else
\language=\csname l@#1\endcsname
\fi
#2}}
\providecommand{\BIBdecl}{\relax}
\BIBdecl

\bibitem{wu2017image}
Q.~{Wu}, C.~{Shen}, P.~{Wang}, A.~{Dick}, and A.~v.~d. {Hengel}, ``Image
  captioning and visual question answering based on attributes and external
  knowledge,'' \emph{IEEE Transactions on Pattern Analysis and Machine
  Intelligence}, vol.~40, no.~06, pp. 1367--1381, 2018.

\bibitem{yang2021unimf}
S.~Yang, R.~Zhang, S.~M. Erfani, and J.~H. Lau, ``Unimf: A unified framework to
  incorporate multimodal knowledge bases intoend-to-end task-oriented dialogue
  systems.'' in \emph{IJCAI}, 2021, pp. 3978--3984.

\bibitem{zhang2016rec}
F.~Zhang, N.~J. Yuan, D.~Lian, X.~Xie, and W.-Y. Ma, ``Collaborative knowledge
  base embedding for recommender systems,'' in \emph{KDD}, 2016, pp. 353--362.

\bibitem{stadler2012linkedgeodata}
C.~Stadler, J.~Lehmann, K.~Hoffner, and S.~Auer, ``Linkedgeodata: A core for a
  web of spatial open data,'' \emph{Semantic Web}, vol.~3, no.~4, pp. 333--354,
  2012.

\bibitem{auer2007dbpedia}
S.~Auer, C.~Bizer, G.~Kobilarov, J.~Lehmann, R.~Cyganiak, and Z.~G. Ives,
  ``Dbpedia: {A} nucleus for a web of open data,'' in \emph{ISWC}, 2007, pp.
  722--735.

\bibitem{zhang2022benchmark}
R.~Zhang, B.~D. Trisedya, M.~Li, Y.~Jiang, and J.~Qi, ``A benchmark and
  comprehensive survey on knowledge graph entity alignment via representation
  learning,'' \emph{The VLDB Journal}, pp. 1--26, 2022.

\bibitem{chen2017multilingual}
M.~{Chen}, Y.~{Tian}, M.~{Yang}, and C.~{Zaniolo}, ``Multilingual knowledge
  graph embeddings for cross-lingual knowledge alignment,'' in \emph{IJCAI},
  2017, pp. 1511--1517.

\bibitem{chen2017multigraph}
M.~{Chen}, T.~{Zhou}, P.~{Zhou}, and C.~{Zaniolo}, ``Multi-graph affinity
  embeddings for multilingual knowledge graphs,'' in \emph{NIPS Workshop on
  Automated Knowledge Base Construction}, 2017.

\bibitem{zhu2017iterative}
H.~{Zhu}, R.~{Xie}, Z.~{Liu}, and M.~{Sun}, ``Iterative entity alignment via
  joint knowledge embeddings,'' in \emph{IJCAI}, 2017, pp. 4258--4264.

\bibitem{cao2019multi}
Y.~Cao, Z.~Liu, C.~Li, Z.~Liu, J.~Li, and T.-S. Chua, ``Multi-channel graph
  neural network for entity alignment,'' in \emph{ACL}, 2019, pp. 1452--1461.

\bibitem{sun2020knowledge}
Z.~Sun, C.~Wang, W.~Hu, M.~Chen, J.~Dai, W.~Zhang, and Y.~Qu, ``Knowledge graph
  alignment network with gated multi-hop neighborhood aggregation,'' in
  \emph{Proceedings of the AAAI Conference on Artificial Intelligence},
  vol.~34, no.~01, 2020, pp. 222--229.

\bibitem{wu2019relation}
Y.~{Wu}, X.~{Liu}, Y.~{Feng}, Z.~{Wang}, R.~{Yan}, and D.~{Zhao},
  ``Relation-aware entity alignment for heterogeneous knowledge graphs,'' in
  \emph{IJCAI}, 2019, pp. 5278--5284.

\bibitem{sun2017cross}
Z.~{Sun}, W.~{Hu}, and C.~{Li}, ``Cross-lingual entity alignment via joint
  attribute-preserving embedding,'' in \emph{ISWC}, 2017, pp. 628--644.

\bibitem{trisedya2019entity}
B.~D. Trisedya, J.~Qi, and R.~Zhang, ``Entity alignment between knowledge
  graphs using attribute embeddings,'' in \emph{Proceedings of the AAAI
  Conference on Artificial Intelligence}, vol.~33, no.~01, 2019, pp. 297--304.

\bibitem{socher2013reasoning}
R.~{Socher}, D.~{Chen}, C.~D. {Manning}, and A.~Y. {Ng}, ``Reasoning with
  neural tensor networks for knowledge base completion,'' in \emph{NIPS}, 2013,
  pp. 926--934.

\bibitem{trisedya2021gcp}
B.~D. Trisedya, J.~Qi, W.~Wang, and R.~Zhang, ``Gcp: Graph encoder with
  content-planning for sentence generation from knowledge bases,''
  \emph{TPAMI}, vol.~44, no.~11, pp. 7521--7533, 2021.

\bibitem{yang2020graphdialog}
S.~Yang, R.~Zhang, and S.~Erfani, ``Graphdialog: Integrating graph knowledge
  into end-to-end task-oriented dialogue systems,'' in \emph{ACL}, 2020.

\bibitem{distiawan2019neural}
B.~D. Trisedya, G.~Weikum, J.~Qi, and R.~Zhang, ``Neural relation extraction
  for knowledge base enrichment,'' in \emph{ACL}, 2019, pp. 229--240.

\bibitem{ding2021prototypical}
N.~Ding, X.~Wang, Y.~Fu, G.~Xu, R.~Wang, P.~Xie, Y.~Shen, F.~Huang, H.-T.
  Zheng, and R.~Zhang, ``Prototypical representation learning for relation
  extraction,'' \emph{ICLR}, 2021.

\bibitem{distiawan2018gtr}
B.~D. Trisedya, J.~Qi, R.~Zhang, and W.~Wang, ``Gtr-lstm: A triple encoder for
  sentence generation from rdf data,'' in \emph{ACL}, 2018, pp. 1627--1637.

\bibitem{trisedya2020sentence}
B.~D. Trisedya, J.~Qi, and R.~Zhang, ``Sentence generation for entity
  description with content-plan attention,'' in \emph{AAAI}, vol.~34, no.~05,
  2020, pp. 9057--9064.

\bibitem{bordes2013translating}
A.~{Bordes}, N.~{Usunier}, A.~{Garcia-Duran}, J.~{Weston}, and O.~{Yakhnenko},
  ``Translating embeddings for modeling multi-relational data,'' in
  \emph{NIPS}, 2013, pp. 2787--2795.

\bibitem{wang2014knowledge}
Z.~{Wang}, J.~{Zhang}, J.~{Feng}, and Z.~{Chen}, ``Knowledge graph embedding by
  translating on hyperplanes,'' in \emph{AAAI}, 2014, pp. 1112--1119.

\bibitem{lin2015learning}
Y.~{Lin}, Z.~{Liu}, M.~{Sun}, Y.~{Liu}, and X.~{Zhu}, ``Learning entity and
  relation embeddings for knowledge graph completion,'' in \emph{AAAI}, 2015,
  pp. 2181--2187.

\bibitem{kipf2017semi}
T.~N. {Kipf} and M.~{Welling}, ``Semi-supervised classification with graph
  convolutional networks,'' in \emph{ICLR}, 2017.

\bibitem{yun2019graph}
S.~Yun, M.~Jeong, R.~Kim, J.~Kang, and H.~J. Kim, ``Graph transformer
  networks,'' \emph{Advances in neural information processing systems},
  vol.~32, 2019.

\bibitem{hu2020heterogeneous}
Z.~Hu, Y.~Dong, K.~Wang, and Y.~Sun, ``Heterogeneous graph transformer,'' in
  \emph{Proceedings of The Web Conference 2020}, 2020, pp. 2704--2710.

\bibitem{mei2022relation}
X.~Mei, X.~Cai, L.~Yang, and N.~Wang, ``Relation-aware heterogeneous graph
  transformer based drug repurposing,'' \emph{Expert Systems with
  Applications}, vol. 190, p. 116165, 2022.

\bibitem{ye2019a}
R.~{Ye}, X.~{Li}, Y.~{Fang}, H.~{Zang}, and M.~{Wang}, ``A vectorized
  relational graph convolutional network for multi-relational network
  alignment.'' in \emph{IJCAI}, 2019, pp. 4135--4141.

\bibitem{liu2020exploring}
Z.~Liu, Y.~Cao, L.~Pan, J.~Li, and T.-S. Chua, ``Exploring and evaluating
  attributes, values, and structures for entity alignment,'' in \emph{EMNLP},
  2020.

\bibitem{li2022uncertainty}
J.~Li and D.~Song, ``Uncertainty-aware pseudo label refinery for entity
  alignment,'' in \emph{The Web Conference}, 2022, pp. 829--837.

\bibitem{gao2021mhgcn}
J.~Gao, X.~Liu, Y.~Chen, and F.~Xiong, ``Mhgcn: Multiview highway graph
  convolutional network for cross-lingual entity alignment,'' \emph{Tsinghua
  Science and Technology}, vol.~27, no.~4, pp. 719--728, 2021.

\bibitem{mangrulkar2022multilingual}
S.~Mangrulkar, A.~MS, and V.~Sembium, ``Multilingual semantic sourcing using
  product images for cross-lingual alignment,'' in \emph{WWW}, 2022, pp.
  41--51.

\bibitem{huang2022multilingual}
Z.~Huang, Z.~Li, H.~Jiang, T.~Cao, H.~Lu, B.~Yin, K.~Subbian, Y.~Sun, and
  W.~Wang, ``Multilingual knowledge graph completion with self-supervised
  adaptive graph alignment,'' in \emph{ACL}, 2022.

\bibitem{ahuja2020language}
A.~Ahuja, N.~Rao, S.~Katariya, K.~Subbian, and C.~K. Reddy, ``Language-agnostic
  representation learning for product search on e-commerce platforms,'' in
  \emph{WSDM}, 2020, pp. 7--15.

\bibitem{zhang2020industry}
Z.~Zhang, J.~Chen, X.~Chen, H.~Liu, Y.~Xiang, B.~Liu, and Y.~Zheng, ``An
  industry evaluation of embedding-based entity alignment,'' in \emph{COLING},
  2020.

\bibitem{liu2023cross}
W.~Liu, J.~Pan, X.~Zhang, X.~Gong, Y.~Ye, X.~Zhao, X.~Wang, K.~Wu, H.~Xiang,
  H.~Yan \emph{et~al.}, ``Cross-platform product matching based on entity
  alignment of knowledge graph with raea model,'' \emph{WWW}, pp. 1--21, 2023.

\bibitem{floridi2020gpt}
L.~Floridi and M.~Chiriatti, ``Gpt-3: Its nature, scope, limits, and
  consequences,'' \emph{Minds and Machines}, vol.~30, pp. 681--694, 2020.

\bibitem{scao2022bloom}
T.~L. Scao, A.~Fan, C.~Akiki, E.~Pavlick, S.~Ili{\'c}, D.~Hesslow,
  R.~Castagn{\'e}, A.~S. Luccioni, F.~Yvon, M.~Gall{\'e} \emph{et~al.},
  ``Bloom: A 176b-parameter open-access multilingual language model,''
  \emph{arXiv preprint arXiv:2211.05100}, 2022.

\bibitem{thoppilan2022lamda}
R.~Thoppilan, D.~De~Freitas, J.~Hall, N.~Shazeer, A.~Kulshreshtha, H.-T. Cheng,
  A.~Jin, T.~Bos, L.~Baker, Y.~Du \emph{et~al.}, ``Lamda: Language models for
  dialog applications,'' \emph{arXiv preprint arXiv:2201.08239}, 2022.

\bibitem{fellbaum1998wordnet}
C.~Fellbaum, \emph{WordNet: An Electronic Lexical Database}.\hskip 1em plus
  0.5em minus 0.4em\relax MIT Press, 1998.

\bibitem{hochreiter1997long}
S.~{Hochreiter} and J.~{Schmidhuber}, ``Long short-term memory,'' \emph{Neural
  Computation}, vol.~9, pp. 1735--1780, 1997.

\bibitem{IPTransE2017}
H.~Zhu, R.~Xie, Z.~Liu, and M.~Sun, ``Iterative entity alignment via joint
  knowledge embeddings.'' in \emph{\textit{IJCAI 2017}}, 2017.

\bibitem{vrandecick2014wikidata}
D.~Vrandecic and M.~Kr{\"{o}}tzsch, ``Wikidata: a free collaborative
  knowledgebase,'' \emph{\textit{CACM}}, vol.~57, no.~10, pp. 78--85, 2014.

\bibitem{hoffart2013yago2}
J.~{Hoffart}, F.~M. {Suchanek}, K.~{Berberich}, and G.~{Weikum}, ``Yago2: A
  spatially and temporally enhanced knowledge base from wikipedia,''
  \emph{Artificial Intelligence}, vol. 194, pp. 28--61, 2013.

\bibitem{sun2019transedge}
Z.~Sun, J.~Huang, W.~Hu, M.~Chen, L.~Guo, and Y.~Qu, ``Transedge: Translating
  relation-contextualized embeddings for knowledge graphs,'' in
  \emph{International Semantic Web Conference}.\hskip 1em plus 0.5em minus
  0.4em\relax Springer, 2019, pp. 612--629.

\bibitem{MultiKE2019}
Q.~Zhang, Z.~Sun, W.~Hu, M.~Chen, L.~Guo, and Y.~Qu, ``Multi-view knowledge
  graph embedding for entity alignment,'' in \emph{\textit{IJCAI}}, 2019.

\bibitem{KECG2019}
C.~Li, Y.~Cao, L.~Hou, J.~Shi, J.~Li, and T.~Chua, ``Semi-supervised entity
  alignment via joint knowledge embedding model and cross-graph model,'' in
  \emph{\textit{EMNLP 2019}}, 2019.

\bibitem{GCN-Align2018}
Z.~Wang, Q.~Lv, X.~Lan, and Y.~Zhang, ``Cross-lingual knowledge graph alignment
  via graph convolutional networks,'' in \emph{\textit{EMNLP 2018}}, 2018.

\bibitem{HGCN2019}
Y.~Wu, X.~Liu, Y.~Feng, Z.~Wang, and D.~Zhao, ``Jointly learning entity and
  relation representations for entity alignment,'' in \emph{\textit{EMNLP
  2019}}, 2019.

\bibitem{GMNN2019}
K.~Xu, L.~Wang, M.~Yu, Y.~Feng, Y.~Song, Z.~Wang, and D.~Yu, ``Cross-lingual
  knowledge graph alignment via graph matching neural network,'' in
  \emph{\textit{ACL 2019}}, 2019.

\bibitem{RDGCN2019}
Y.~Wu, X.~Liu, Y.~Feng, Z.~Wang, R.~Yan, and D.~Zhao, ``Relation-aware entity
  alignment for heterogeneous knowledge graphs,'' in \emph{\textit{IJCAI
  2019}}, 2019.

\bibitem{CEA2020}
W.~Zeng, X.~Zhao, J.~Tang, and X.~Lin, ``Collective entity alignment via
  adaptive features,'' in \emph{\textit{ICDE 2020}}, 2020.

\bibitem{MRAEA2020}
X.~Mao, W.~Wang, H.~Xu, M.~Lan, and Y.~Wu, ``{MRAEA:} an efficient and robust
  entity alignment approach for cross-lingual knowledge graph,'' in
  \emph{\textit{WSDM 2020}}, 2020.

\bibitem{NMN2020}
Y.~Wu, X.~Liu, Y.~Feng, Z.~Wang, and D.~Zhao, ``Neighborhood matching network
  for entity alignment,'' in \emph{\textit{ACL 2020}}, 2020.

\bibitem{sun2020benchmarking}
Z.~Sun, Q.~Zhang, W.~Hu, C.~Wang, M.~Chen, F.~Akrami, and C.~Li, ``A
  benchmarking study of embedding-based entity alignment for knowledge
  graphs,'' in \emph{\textit{VLDB 2020}}, 2020.

\bibitem{jiang2018hyperx}
W.~Jiang, J.~Qi, J.~X. Yu, J.~Huang, and R.~Zhang, ``Hyperx: A scalable
  hypergraph framework,'' \emph{TKDE}, vol.~31, no.~5, pp. 909--922, 2018.

\bibitem{su2021detecting}
Y.~Su, R.~Zhang, S.~Erfani, and Z.~Xu, ``Detecting beneficial feature
  interactions for recommender systems,'' in \emph{AAAI}, 2021, pp. 4357--4365.

\bibitem{su2022detecting}
Y.~Su, Y.~Zhao, S.~Erfani, J.~Gan, and R.~Zhang, ``Detecting arbitrary order
  beneficial feature interactions for recommender systems,'' in \emph{KDD},
  2022, pp. 1676--1686.

\bibitem{zhao2023learning}
Y.~Zhao, J.~Qi, B.~D. Trisedya, Y.~Su, R.~Zhang, and H.~Ren, ``Learning region
  similarities via graph-based deep metric learning,'' \emph{TKDE}, 2023.

\end{thebibliography}
\vspace{-1cm}

\begin{IEEEbiography}[{\includegraphics[width=1in,height=1in,clip,keepaspectratio]{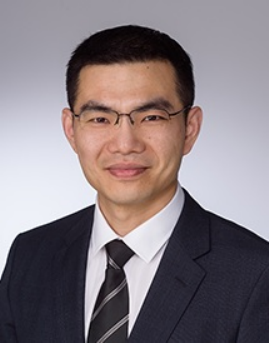}}]{Rui Zhang}
is a visiting Professor at Tsinghua University. His research interests include big data, data mining, and machine learning. Professor Zhang has won several awards, including Future Fellowship by the Australian Research Council in 2012, Chris Wallace Award for Outstanding Research by the Computing Research and Education Association of Australasia in 2015, and Google Faculty Research Award in 2017.
\end{IEEEbiography}
\vspace{-1.4cm}
\begin{IEEEbiography}[{\includegraphics[width=1.0in,height=0.9in,clip,keepaspectratio]{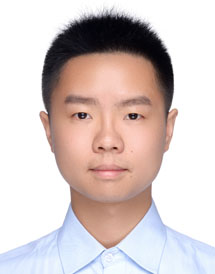}}] 
{Yixin Su} received his Master's and Ph.D. degrees from the School of Computing and Information Systems at The University of Melbourne. His research interests include leveraging graph neural networks to enhance feature interaction modeling and collaborating filtering in recommender systems.
\end{IEEEbiography}
\vspace{-1.5cm}
\begin{IEEEbiography}[{\includegraphics[width=1in,height=1.1in,clip,keepaspectratio]{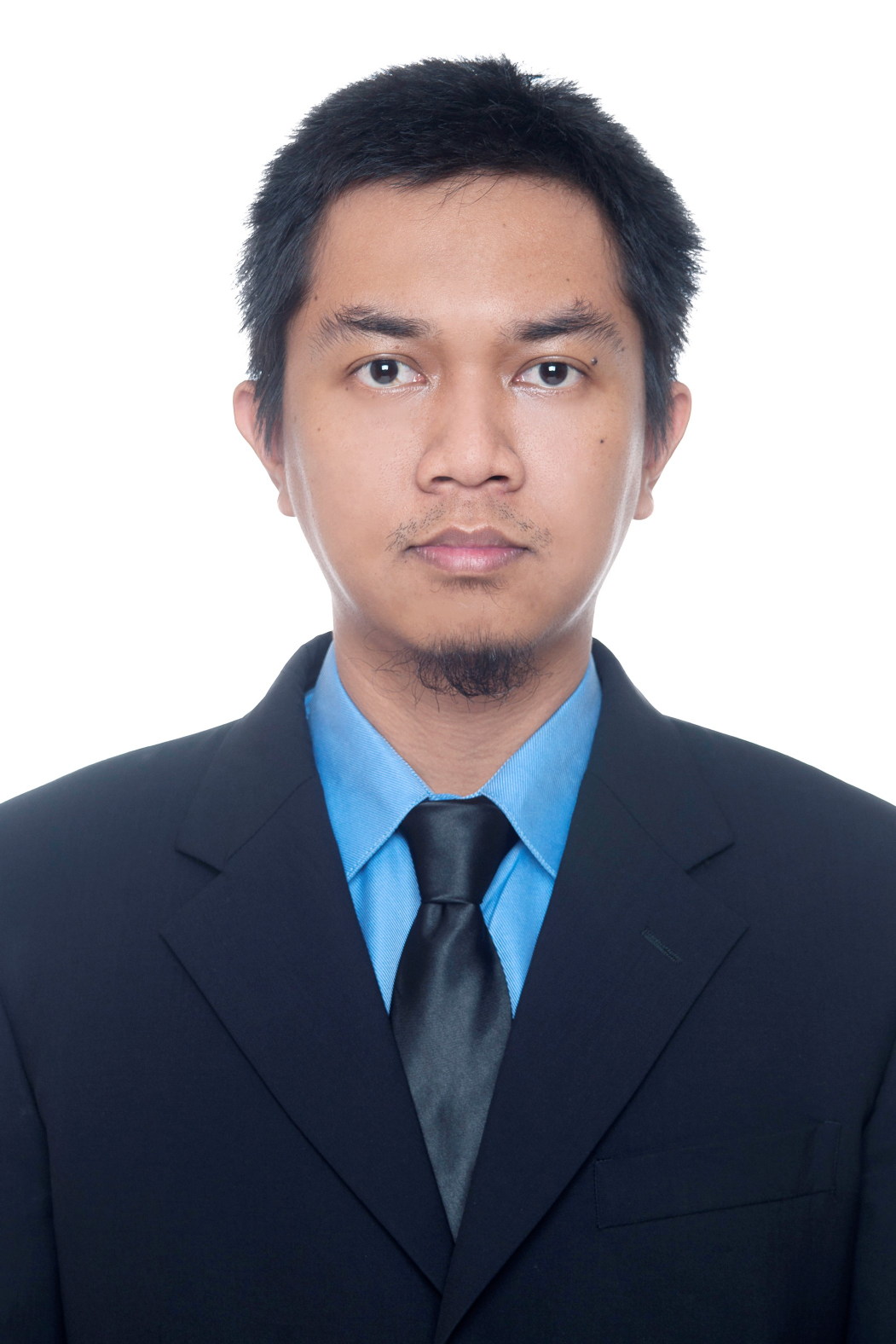}}]{Bayu Distiawan Trisedya}
is a Lecturer in the Faculty of Computer Science Universitas Indonesia. He was a Postdoctoral Research Fellow in the School of Computing and Information Systems at The University of Melbourne and RMIT University. He received his bachelor’s and Master’s degrees from Universitas Indonesia in 2009 and 2011, respectively. He received his Ph.D. degree from The University of Melbourne in 2021. His research interest is knowledge graphs and natural language processing.
\end{IEEEbiography}
\vspace{-1.1cm}
\begin{IEEEbiography}[{\includegraphics[width=1.0in,height=0.9in,clip,keepaspectratio]{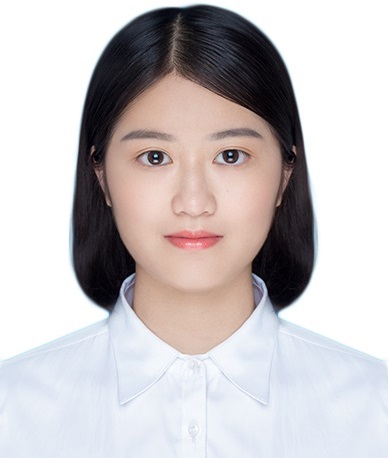}}] 
{Xiaoyan Zhao} is currently a PhD student with the Department of Systems Engineering and Engineering Management, the Chinese University of Hong Kong. She received the B.S. degree from Wuhan University of Technology in 2019, and M.E. degree from University of Chinese Academy of Sciences in 2022. Her research interests include natural language processing and knowledge graph.
\end{IEEEbiography}
\vspace{-1.2cm}
\begin{IEEEbiography}[{\includegraphics[width=1.0in,height=0.9in,clip,keepaspectratio]{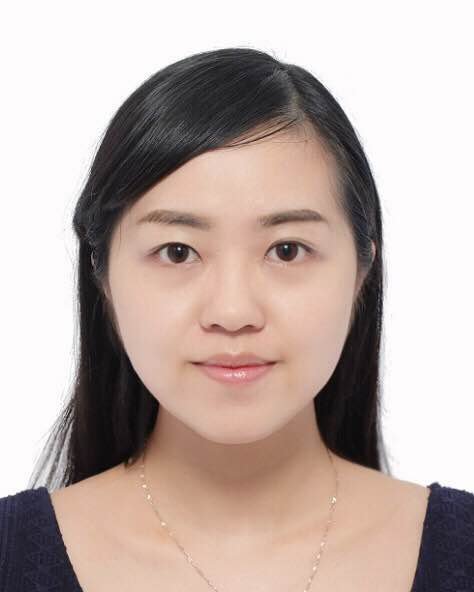}}]
{Min Yang} is currently an Associate Professor with the Shenzhen Institutes of Advanced Technology, Chinese Academy of Science. She received her Ph.D. degree from the University of Hong Kong in February 2017. Prior to that, she received her B.S. degree from Sichuan University in 2012.  Her current research interests include machine learning, deep learning and natural language processing.
\end{IEEEbiography}
\vspace{-1.2cm}
\begin{IEEEbiography}[{\includegraphics[width=1.0in,height=0.9in,clip,keepaspectratio]{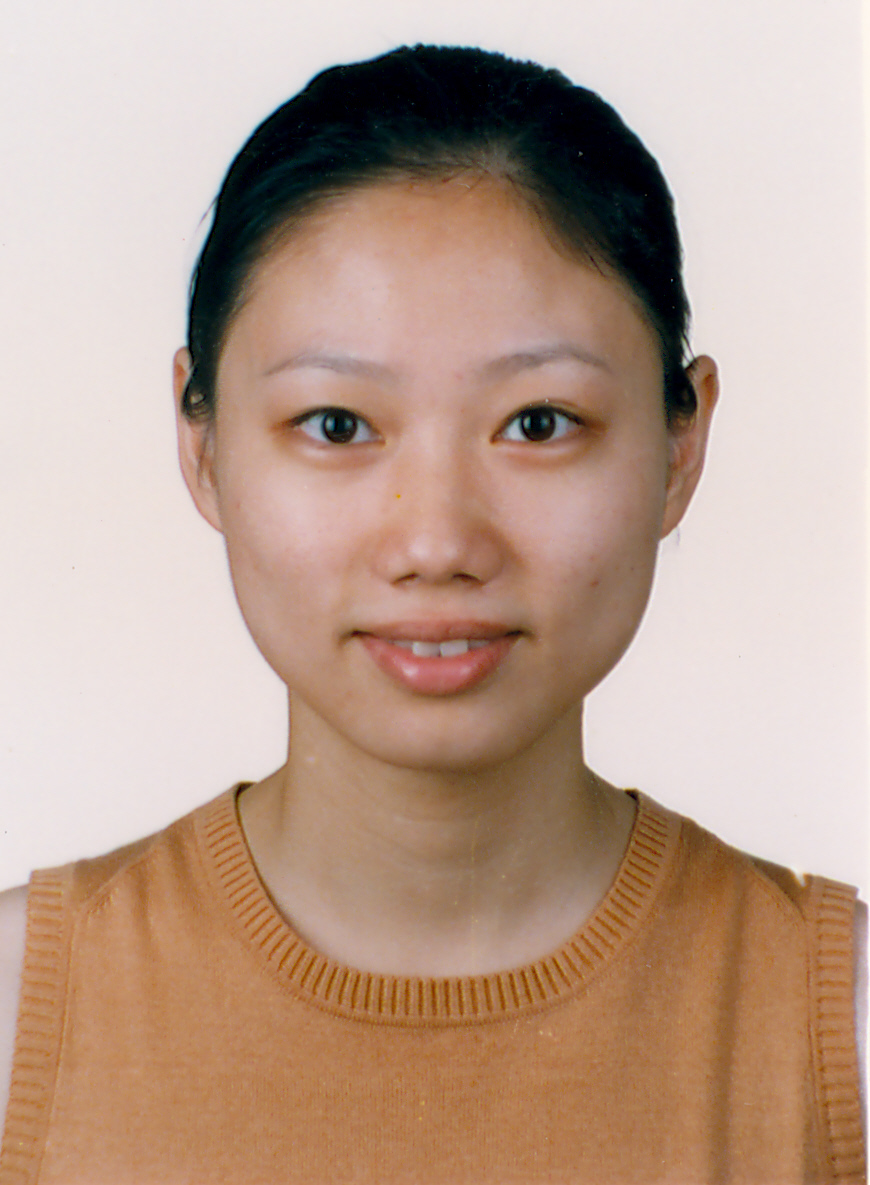}}]
{Hong Cheng} is a Professor in the Department of Systems Engineering and Engineering Management at the Chinese University of Hong Kong. She received her Ph.D. degree from University of Illinois at Urbana-Champaign in 2008. Her research interests include data mining, database systems, and machine learning. She received research paper awards at ICDE’07, SIGKDD’06 and SIGKDD’05, and the certificate of recognition for the 2009 SIGKDD Doctoral Dissertation Award. She is a recipient of the 2010 Vice-Chancellor’s Exemplary Teaching Award at the Chinese University of Hong Kong.
\end{IEEEbiography}
\vspace{-1cm}
\begin{IEEEbiography}[{\includegraphics[width=1in,height=1.1in,clip,keepaspectratio]{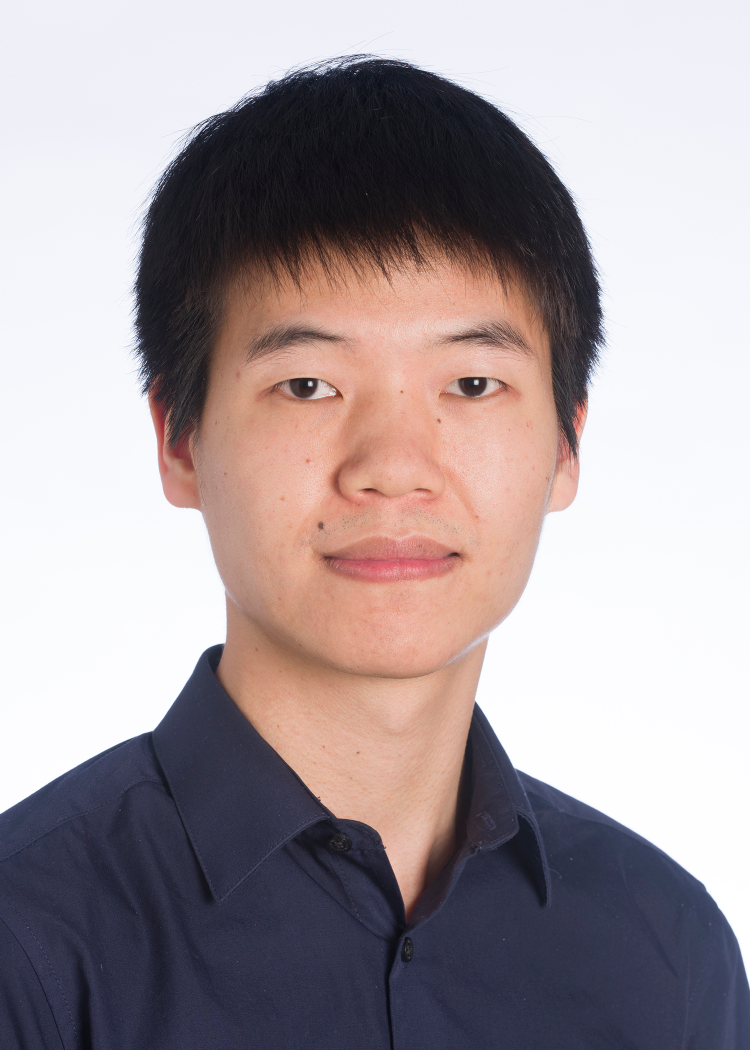}}]{Jianzhong Qi}
is a Senior Lecturer in the School of Computing and Information Systems at The University of Melbourne. He received his Ph.D. degree from The University of Melbourne in 2014. His research interests include machine learning and data management and analytics, with a focus on spatial, temporal, and textual data.
\end{IEEEbiography}
\vfill

\newpage
\clearpage
\setcounter{page}{1}
\color{black}
\appendices
\section{Automatically Obain Entity Types}
\label{appx:obtain_type}

To construct the predicate-proximity-graph, we automatically obtain the types of each entity by extracting them from the SPARQL Query Editor (\url{https://dbpedia.org/sparql}). Specifically, we obtain the types of entities through the following steps:

\begin{itemize}
\item For each entity, e.g., Barack Obama, we convert it into a DBpedia graph dataset format, e.g., \url{http://dbpedia.org/resource/Barack_Obama}.
\item Then, we search for the types of the entity through the query:

\lstset{
  language=SQL,
  basicstyle=\ttfamily,
  keywordstyle=\bfseries,
  commentstyle=\itshape,
  showstringspaces=false,
  columns=flexible,
  frame=single,
  breaklines=true,
  postbreak=\mbox{\textcolor{red}{$\hookrightarrow$}\space},
}

\begin{lstlisting}
PREFIX rdf: <http://www.w3.org/1999/02/22-rdf-syntax-ns#>
PREFIX rdfs: <http://www.w3.org/2000/01/rdf-schema#>
PREFIX dbr: <http://dbpedia.org/resource>
PREFIX dbo: <http://dbpedia.org/ontology>
SELECT DISTINCT ?obj WHERE {
<http://dbpedia.org/resource/Barack_Obama> rdf:type ?obj
FILTER strstarts(str(?obj), str(dbo:))
}
\end{lstlisting}

The first four lines define the prefixes for the namespaces used in the query. ``rdf:" and ``rdfs:" are standard namespaces for RDF and RDF Schema, respectively. "dbr:" and "dbo:" are prefixes for the DBpedia resource and ontology namespaces, respectively; the SELECT clause specifies that we want to retrieve the distinct values of the variable "?obj", which represents the types of which "Barack Obama" is an entity.
\item Finally, we get a set of types that belongs to the entity, e.g., {Person, Politician, OfficeHolder}, which will replace the entity "Barack Obama" in the predicate-proximity-graph.
\item If we cannot obtain any type from the entity, we will keep the entity as it is.
\end{itemize}

\color{black}

\end{document}